\documentclass[10pt,aps,prf,showpacs,onecolumn,superscriptaddress,floats,floatfix,longbibliography]{revtex4-2}

\usepackage{graphicx}
\usepackage{dcolumn}
\usepackage{bm}
\usepackage{hyperref}
\usepackage{color}
\usepackage{natbib}
\usepackage{amsmath} 
\usepackage{amssymb}
\usepackage{booktabs}
\usepackage{array}
\usepackage{tabularx}
\usepackage[normalem]{ulem}
\newcolumntype{C}{>{\centering\arraybackslash}X}

\newcommand{\vertbar}{\rule[-1ex]{0.5pt}{2.5ex}}
\begin{document}

\title{Interface Fluctuations in a Turbulent Binary Fluid using Data-Driven Methods}

\author{Samuel Z Khiangte}
\affiliation{Center for Condensed Matter Theory, Indian Institute of Science, Bangalore, India}

\author{Triparna Sanyal}
\affiliation{Centre for Ocean, River, Atmosphere and Land Sciences, Indian Institute of Technology, Kharagpur, India}

\author{Sumantra Sarkar}
\affiliation{Center for Condensed Matter Theory, Indian Institute of Science, Bangalore, India}
\author{Nairita Pal}
\affiliation{Centre for Ocean, River, Atmosphere and Land Sciences, Indian Institute of Technology, Kharagpur, India}

\begin{abstract}

Interfacial fluctuations in a two-phase binary fluid mixture reveal signatures of underlying physical processes that occur within each phase and on a range of spatial and temporal scales. In this study, we investigate a model binary fluid system consisting of a single droplet of one phase moving in the background of the second phase. The binary fluid system is subjected to turbulent forcing. We perform extensive direct numerical simulations of the turbulent system to examine how quantities such as interfacial dynamics and droplet acceleration can be systematically decoded. Extensive simulations of binary fluid systems are computationally expensive and time-consuming. In contrast, data-driven models have shown promise in recent times in reducing computational cost. In this work, we build and compare the performances of four interpretable data-driven models, i.e., dynamic mode decomposition (DMD), Hankel DMD, sparse identification of nonlinear dynamics (SINDy), and Stochastic Langevin regression (SLR),  each using dimensionality reduction via proper orthogonal decomposition, to identify simplified dynamical equations governing interfacial dynamics and center-of-mass acceleration. We show how these learned models can be generalized to encode physical properties, such as the interfacial surface tension and droplet size. In particular, we show that SLR predicts the underlying dynamical equations of the binary-fluid system with the greatest accuracy over a wide range of interfacial tension values and droplet sizes. In addition, SLR requires fewer terms compared to SINDy to capture the underlying dynamics, and is thus computationally the most efficient among the four methods. These data-driven techniques can be used in many practical applications, such as the dynamics of biological cell membranes, thin films, and other industrial applications.

\end{abstract}

\date{\today}

\maketitle


\section{\label{sec:level1}Introduction}
The study of finite-sized deformable droplets advected by turbulent flows~\cite{Pope_2000} forms the basis of multiphase flows~\cite{brennen2005fundamentals1}, which is an active
area of research in several industrial and natural settings. Droplet advection by underlying fluids has wide applications ranging from analysis of oil spills in oceans~\cite{chen2018numerical,boufadel2020transport}, dispersion of pollutants and toxic elements in the ocean \cite{chen2018numerical, sutherland2023fluid} and the atmosphere \cite{kumar2019optimum,qureshi2007turbulent}, formation of clouds~\cite{shaw2003particle, chandrakar2023lagrangian}, analysis of volcanic plumes, and spread of diseases such as COVID-19 through infected water droplets in human breath \cite{bourouiba2021fluida,bourouiba2021fluidb}. Droplet advection in background flows has been extensively studied in both laminar and turbulent regimes~\cite{badalassi2003computation, roccon2023phase, elghobashi2019direct,bec2024statistical,balachandar2010turbulent, sommerfeld2017numerical, riviere2021sub,homann2010finite,speetjens2021lagrangian,pal2016binary,pal2022ephemeral,albernaz2017droplet}. Such studies have discovered interesting phenomena such as droplet shape fluctuations~\cite{pal2016binary,albernaz2017droplet} and droplet center-of-mass motion under the effect of a background turbulent flow. Time-dependent fluctuations in the shape of a droplet moving in a turbulent background fluid occur due to interfacial surface tension forces, as well as the imposed background turbulent forcing. These fluctuations have been shown to have a multifractal nature in \cite{pal2016binary}. In addition, the background turbulent force also causes motion of the droplet center of mass. Acceleration of the center of mass of the droplets has been shown to exhibit strong intermittency as the size of the droplets reduces~\cite{pal2016binary,homann2010finite}. Such droplet dynamics phenomena also aid in the study of droplet or particle clustering when studying the formation of clouds \cite{fedorets2017self,gumber2024turbulence}.

\medskip

The droplet coupled with the background flow can be characterized as a multiphase system, where the droplet constitutes one phase, and the background fluid constitutes the second phase. Finite-sized and deformable droplets affect the background fluid considerably as they are transported by the background flow. Thus, in addition to the droplet transport, we have to take into account the time--dependent deformations of the droplet boundary. Various numerical schemes have been adopted to execute the high-fidelity numerical simulations of the advection of droplets in a background fluid. Examples include widely used phase-field models~\cite{badalassi2003computation, roccon2023phase, pal2016binary}, the moving boundary methods~\cite{balachandar2010turbulent, sommerfeld2017numerical}, and Lattice-Boltzmann methods~\cite{biferale2012convection}. In our work, we use the phase--field model to solve the multiphase flow problem in two spatial dimensions (2D). Simulations based on phase-field models track the moving interface between the droplet and the background fluid through gradients of an order parameter field $\phi$. Thus, tracking of the deforming boundary between the droplet and fluid becomes seamless and computationally cheap in the phase--field model. In particular, we use the Cahn--Hilliard--Navier--Stokes (CHNS) equations~\cite{pal2016binary,pal2022ephemeral} to simulate the droplet motion via gradients in order--parameter field $\phi$. The CHNS equations have been used extensively in studies of phase transitions~\cite{boffetta2012two}, nucleation~\cite{lothe1962reconsiderations}, and spinodal decomposition~\cite{onuki2002phase}. We use the CHNS equations to carry out a detailed DNS study of droplet dynamics in turbulent flow and characterize the turbulence-induced droplet shape deformations and their acceleration statistics.


\medskip

The drawback of high-fidelity numerical simulations and experiments is that they are time-consuming and expensive. For example, the DNS of the CHNS equations presented in this paper required approximately 15 days on GPUs. Thus, it becomes pertinent to build simpler models to capture the dynamics of multiphase flows more efficiently and using fewer resources than high-fidelity numerical simulations. In this work, we investigate the efficacy of recent advances in interpretable data-driven methods in extracting such reduced, physically transparent models directly from time-series data, targeting two key quantities: (1) the droplet interface dynamics and (2) the droplet acceleration. For spatiotemporal data such as droplet interface dynamics, Proper Orthogonal Decomposition (POD) can be used to represent the system in an orthogonal basis that captures the dominant modes of deformation, thereby reducing the dimensionality and complexity of the dynamics by truncating less significant modes.
Here, we combine POD with four data-driven techniques to model droplet dynamics and validate the results with DNS. First, we use the Dynamic Mode Decomposition (DMD)~\cite{schmid:hal-01020654, schmid2022dynamic, kutz2016dynamic,PhysRevE.104.015206} technique, which is a dimensionality reduction method used to identify dominant spatio-temporal patterns in time-dependent data. Although DMD can be applied to non-linear systems, the primary drawback of DMD is that it employs a linear operator to approximate the dynamics of the system under study. Hence, as we demonstrate in our paper, DMD is unable to capture and forecast the time-dependent evolution of droplet shape fluctuations and the droplet motion as a whole, even with improved methods such as the Hankel DMD. In Section~\ref{sec:level4}B, we show how the DMD-based methods fail to predict the turbulent dynamics. 

DMD-based methods use a linear operator to model a nonlinear system. To address this problem, in Section~\ref{sec:level3}C, we use the Sparse Identification of Nonlinear Dynamics (SINDy), a data-driven method to discover nonlinear governing equations of dynamical systems from simulation or experimental data ~\cite{doi:10.1073/pnas.1517384113, fukami2021sparse}. As we show in Section~\ref{sec:level4} of our paper, SINDy can capture and forecast the droplet interface fluctuation for a given surface tension, but the learned models do not generalize well to other surface tensions. A more effective approach to modeling the time-dependent dynamics of droplets is to employ a stochastic method coupled with SINDy. Many physical systems characterized by nonlinear multiscale interactions can be modeled by considering the unresolved degrees of freedom as random fluctuations \cite{Risken1996,Friedrich2011}. However, turbulent systems do not always have merely random fluctuations and need to be treated via stochastic and non-Gaussian approaches ~\cite{Landau1987Fluid, callaham2021nonlinear}. In our work, we refer to this method as stochastic Langevin regression \cite{callaham2021nonlinear}. Among the four data-driven methods, Stochastic Langevin Regression (SLR) agrees best with the DNS data. While DMD, SINDy, and stochastic regression have individually been applied to fluid dynamics problems~\cite{schmid2022dynamic, doi:10.1073/pnas.1517384113, callaham2021nonlinear}, the present work makes several distinct contributions. First, we provide a systematic comparison of these four methods applied to the same physical system, enabling a controlled assessment of their relative merits for modeling turbulent multiphase flows. Second, we introduce a physically motivated generalization procedure based on the observation that POD singular values encode the dependence on surface tension, enabling trained models to predict dynamics at untrained parameter values. Third, we demonstrate that stochastic models require significantly fewer modes than deterministic models to capture the deformation statistics, providing insight into the role of unresolved modes as effective noise sources. Fourth, for the center-of-mass acceleration, we discover stochastic equations whose coefficients vary systematically with droplet size, extending previous results for tracer particles~\cite{PhysRevFluids.4.044301} to finite-size deformable droplets.

The remainder of the paper is organized as follows. Section~\ref{sec:level3}A introduces the DNS equations and the methods used to solve them. In Section~\ref{sec:level3}B, we discuss an overview of the data-driven methods and present a flowchart for the model discovery steps we follow. 
Section~\ref{sec:level4} discusses the results and analyzes how the above-mentioned methods can be used to learn equations for the droplet interface dynamics. 
Section~\ref{sec:level4}A introduces the Proper Orthogonal Decomposition and discusses how it can be applied to reduce model complexity. Section~\ref{sec:level4}B introduces the DMD techniques used in our paper, and the numerical methods used for the implementation, along with an analysis of the performance of the learned models in predicting the interface dynamics. Section~\ref{sec:level4}C introduces the SINDy method, and also the numerical methods used for implementing it, along with an analysis of the performance of the predicted models. In Section~\ref{sec:level4}D, we introduce the Stochastic Langevin Regression method, which uses parts of the SINDy method to predict stochastic differential equations. Section~\ref{sec:level4}E discusses how the same methods can be used to learn models for the droplet data and analyzes the models' performance when compared to DNS results. In this section, we also show how changing certain coefficients that affect the acceleration dynamics, like the droplet size, can affect the coefficients of the learned model. 
Finally, we present the discussion of our results and conclusion in Section~\ref{conclusion}.



\section{\label{sec:level3}Methods}
\subsection{Direct Numerical Simulation}
For a model system, we use the Cahn-Hilliard Navier-Stokes (CHNS) equation written in terms of the vorticity field and solve the equations using the procedure prescribed in \cite{pal2016binary}. The full model system is described below. 

\begin{align}
    (\partial_t + \mathbf{u} \cdot \nabla)\omega &= \nu \nabla^2 \omega - \alpha \omega - \nabla \times (\phi \nabla n) + F_\omega, \label{eq:omega} \\
    (\partial_t + \mathbf{u} \cdot \nabla)\phi &= \gamma \nabla^2 n, \label{eq:phi} \\
    \nabla \cdot \mathbf{u} &= 0. \label{eq:incompressible}
\end{align}
\begin{align}
    \omega &= (\nabla \times \mathbf{u}) \cdot \hat{\mathbf{e}}_z, \label{eq:vorticity_def} \\
    n(\mathbf{x}, t) &= \frac{\delta \mathcal{F}[\phi]}{\delta \phi}, \label{eq:chem_pot} \\
    \mathcal{F}[\phi] &= \Lambda \int \left[ \frac{(\phi^2 - 1)^2}{4\xi^2} + \frac{|\nabla \phi|^2}{2} \right] d\mathbf{x}. \label{eq:free_energy}
\end{align}
In Eq.~\eqref{eq:omega} ${\mathbf u}$ is the fluid velocity, $\omega$ as defined in Eq.~\eqref{eq:vorticity_def} is the fluid vorticity, $\nu$ is the kinematic viscosity, $\alpha$ is the coefficient of friction induced by surface drag with air, $n$ as defined in Eq.~\eqref{eq:chem_pot} is a chemical potential, $F_\omega = F_0 \cos(k_f y)$ in Eq.~\eqref{eq:omega} is a Kolmogorov-type forcing which drives the system~\cite{CHILDRESS2001105}, with $F_0$ representing the forcing amplitude and $k_f$ the forcing wavenumber. ${\mathbf u}$ follows the incompressibility equation \eqref{eq:incompressible}. ${\mathcal F}$ as defined in Eq.~\eqref{eq:free_energy} is a free energy functional. In Eq.~\eqref{eq:free_energy} $\Lambda$ is the density of energy with which the two phases mix at the interface, $\xi$ sets the scale of the interfacial width of the binary fluid mixture. In Eq.~\eqref{eq:phi} $\gamma$ is the coefficient of mobility of the binary fluid mixture. In our work, the surface tension is given by 
\begin{equation}
    \sigma = \displaystyle\frac{2\sqrt{2}\Lambda}{3\xi}
    \label{surface_tension}
\end{equation}
 and the Weber number defined as $\displaystyle\frac{\rho L_f^3 F_0}{\sigma}$ (here $L_f=\displaystyle\frac{2\pi}{k_f}$) gives a natural dimensionless measure of the surface tension. 

In our work, $\phi(x,t) > 0$ determines the background (majority) phase and $\phi(x,t) < 0$ determines the droplet (minority) phase of our binary fluid mixture. We initialize the phase field $\phi(x, y)$ with a circular membrane profile at $t=0$, given by:
\begin{equation}
    \phi(x, y) = \tanh\left\{
    \frac{1}{\sqrt{2}\xi}
    \left[
        \sqrt{(x - x_c)^2 + (y - y_c)^2} - \frac{d_0}{2}
    \right]
    \right\},\label{initialize}
\end{equation}
where $(x_c, y_c)$ denotes the circle center, $d_0$ is the initial diameter, and $\xi$ determines the width of the diffuse interface. The interface width is made non-dimensional with the Cahn number $Ch=\displaystyle\frac{\xi}{L}$, where $L$ is the length of the simulation domain. The initialization of $\phi(x,y)$ in Eq.~\eqref{initialize} ensures that the droplet is circular at $t=0$. 

Our direct numerical simulations (DNSs) of Eqs.~\eqref{eq:omega}
and \eqref{eq:phi} are solved using a pseudo-spectral method and periodic boundary
conditions for spatial derivatives; $L(= 2\pi)$ is the size of our square simulation
domain, which has $N^2$ collocation points. We have a cubic
nonlinearity in $\phi$ in Eq.~\eqref{eq:phi}, which is due to the form of the chemical potential $n$ [follows from Eq.~\eqref{eq:chem_pot}]. Due to this cubic non-linearity, we
use $N/2$--dealiasing \cite{canuto2007spectral}. For time integration of Eqs.~\eqref{eq:omega} and \eqref{eq:phi} we use the
exponential Adams-Bashforth method ETD2~\cite{cox2002exponential}. We use
computers with Graphics Processing Units (e.g., the NVIDIA 3070Ti), which we program in CUDA~\cite{cudasite}. Due to our efficient CUDA programming, we are able to explore a large parameter space of the CHNS equations, and also carry out long-time simulations essential for our studies. In the following paragraph we introduce the mean quantities such as the droplet deformation parameter and the center of mass acceleration. We calculate these quantities from the field variables $\omega(x,t)$ and $\phi(x,t)$ obtained from DNSs of Eqs.~\eqref{eq:omega} and \eqref{eq:phi}. The evolution of the system through time is shown in Figure~\ref{schematic}(A-C), where the interface of the phases is marked in red.



\subsubsection{Definition of Variables}

The dynamical quantities analyzed in this study are defined as follows, in accordance with the framework introduced in Ref.~\cite{pal2016binary}. First, the deformation of the droplet interface is characterized by the instantaneous shape metric $\Gamma(t)$, defined via the ratio of the instantaneous perimeter $P(t)$ to the initial perimeter of an undeformed circular droplet of the same area $P_0(t)$:
\begin{equation}
\Gamma(t) = \frac{P(t)}{P_0(t)}-1 ;
\label{deform}
\end{equation}

\begin{figure}[h!]
\centering
\includegraphics[scale=0.11]{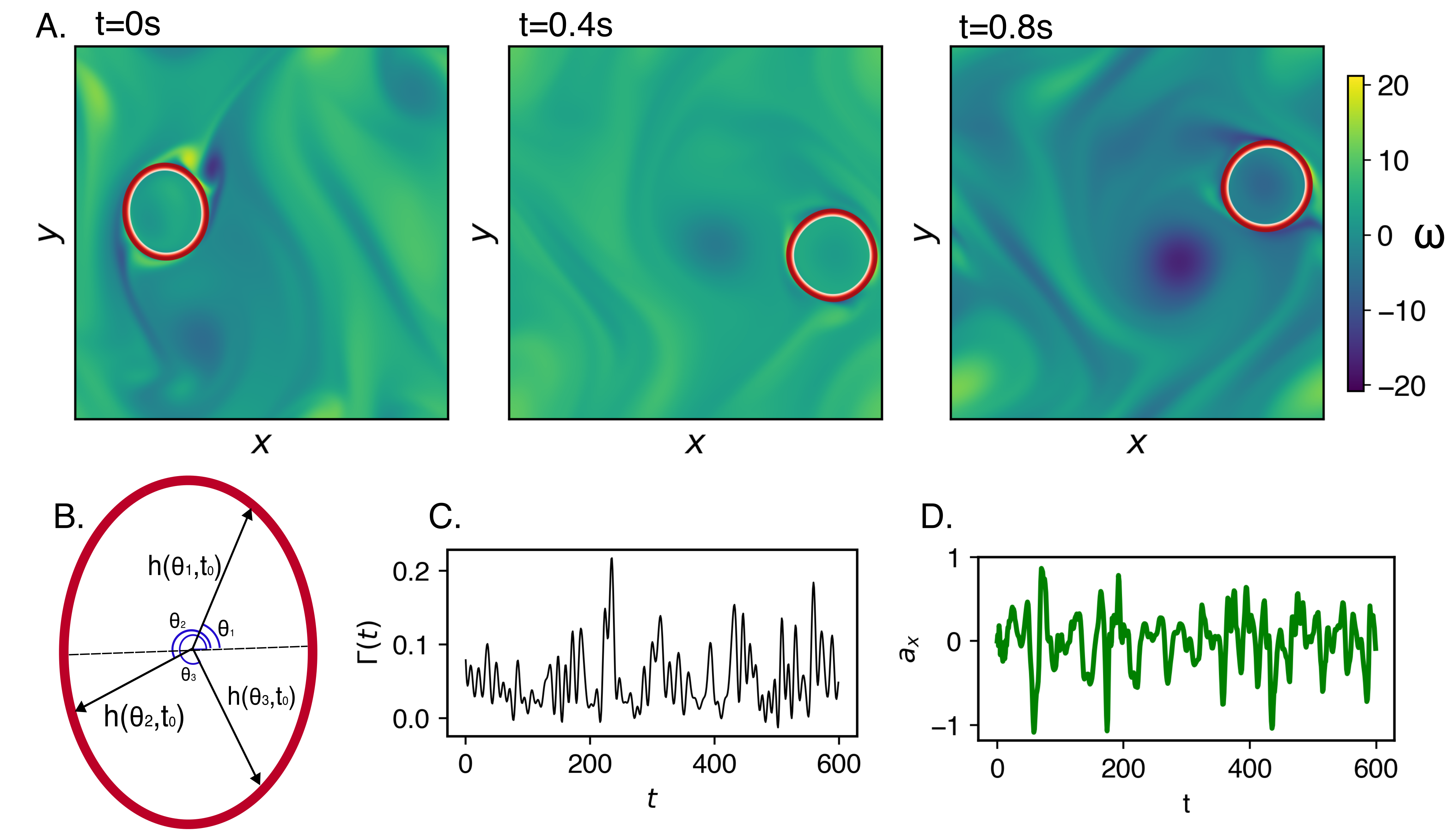}
\caption{\textbf{Extracting measurable droplet properties from DNS:} (A) Droplet under the influence of turbulent flow with a surface tension corresponding to $\Lambda=200$. The underlying flow is color-coded by the vorticity field, with red denoting the droplet interface. (B) Definition of height ($h_i$) field at different $\theta_i$ at a given time $t_0$.
(C) Deformation parameter calculated from DNS. (D) Acceleration calculated from DNS.}
\label{schematic}
\end{figure}

$P(t)$ and $P(0)$ can be computed numerically using the computational points on the droplet surface. We define the instantaneous droplet radius vector for the $i$‑th surface point as $\mathbf{h}_i = h_{x,i}\hat{\mathbf{x}} + h_{y,i}\hat{\mathbf{y}} = h_i \cos\theta_i\,\hat{\mathbf{x}} + h_i \sin\theta_i\,\hat{\mathbf{y}}$, where $h_i = \sqrt{h_{x,i}^2 + h_{y,i}^2}$ is the distance of that surface point from the droplet’s center of mass, and $\theta_i = \tan^{-1}(h_{y,i}/h_{x,i})$ is the angle the radius vector makes with the $X$-axis (see Figure~\ref{schematic}(B)).

For a perfectly spherical droplet, as we move along the perimeter, there is a change in $\theta$ only, and $h$ remains unchanged. For a deformed droplet, there is a change in $\theta$, as well as $h$. Numerically, $P(t)$ can be computed as
\begin{equation}
P(t) = \sum_{i=0}^{N} \sqrt{ [h_{i+1}(t) - h_i(t)]^2 + [h_i(t) \cdot (\theta_{i+1}(t) - \theta_i(t))]^2 }.
\label{peri_calc}
\end{equation}
Here $N$ is the set of points lying on the droplet surface. 
For this work, we discretize the angular space $\theta$ into $N=150$ points between $0$ and $2\pi$ and calculate the radius $h_i$ at each of the discretized angles $\theta_i$.



\medskip

Second, the translational motion of the droplet is described by the center of mass (COM) velocity and acceleration fields. The instantaneous COM velocity $\mathbf{v}_{\text{CM}}(t)$ is obtained by spatially averaging the velocity field $\mathbf{u}(\mathbf{x},t)$ over the droplet interior, defined as the region where $\phi(\mathbf{x},t) < 0$:
$
\mathbf{v}_{\text{CM}}(t) = \frac{1}{N_{\text{drop}}} \sum_{\substack{\mathbf{x} \ \phi(\mathbf{x},t) < 0}} \mathbf{u}(\mathbf{x},t),
$
where $N_{\text{drop}}$ is the number of grid points in the droplet interior.
Correspondingly, the components of the COM acceleration, $a_x(t)$ and $a_y(t)$, are computed as the time derivative of the COM velocity:

\begin{equation}
a_i(t) = \frac{d v_{\text{CM},i}(t)}{dt}
\end{equation}
In Section~\ref{Generalisation}, we will show how the extracted models can be extrapolated to predict interfacial dynamics at different surface tensions. The surface tension is varied by changing the energy density ($\Lambda$) associated with the mixing of the two phases at the interface (Eq.~\eqref{surface_tension}).

\noindent To analyze the interfacial dynamics, we construct a time-resolved height matrix $H(t)$, which encapsulates the interface profile $h_\theta(t_i)$ at each angular coordinate $\theta_i$ and time point $t_i$ (Figure~\ref{schematic}(A)). The coordinates of the droplet interface are divided into $N=150$ equally spaced points. We take training data for $T_{train}=7000$ steps and test the result for $T_{sim}=800$ steps, where each time step corresponds to $dt=0.0002$\,s.
\begin{equation}
H(\mathbf{\theta},\mathbf{t}) =
\begin{bmatrix}
h(\theta_1,t_1) & h(\theta_1,t_2) & \cdots & h(\theta_1,t_T) \\
h(\theta_2,t_1) & h(\theta_2,t_2) & \cdots & h(\theta_2,t_T) \\
\vdots   & \vdots   & \ddots & \vdots   \\
h(\theta_N,t_1) & h(\theta_N,t_2) & \cdots & h(\theta_N,t_T)
\end{bmatrix}
\label{H_matrix}
\end{equation}

{\subsection{Predictive modeling using interpretable data-driven methods}
In this manuscript, we use data-driven methods to find simple, effective descriptions of droplet dynamics. In principle, it is possible to integrate out the fast degrees of freedom in the CHNS equations to write down an effective model for the droplet dynamics. However, the complete dynamics of droplets in turbulent flows is complex and involves processes happening at multiple scales. Hence, identifying the fast and integrable degrees of freedom is nontrivial. Data-driven methods have been successful in identifying such reduced-order effective models in various situations. Hence, in this paper, we use this approach to derive effective models for the droplet dynamics using three interconnected approaches. Specifically, we identify an effective model of droplet fluctuation in the center of mass frame of the droplets and another model to describe the dynamics of the center of mass. In the 'Results' section and in the 'Appendix', we provide a detailed description of these methods. However, for the sake of clarity, we describe them briefly here. 

\begin{figure}[h!]
\centering
\includegraphics[scale=0.11]{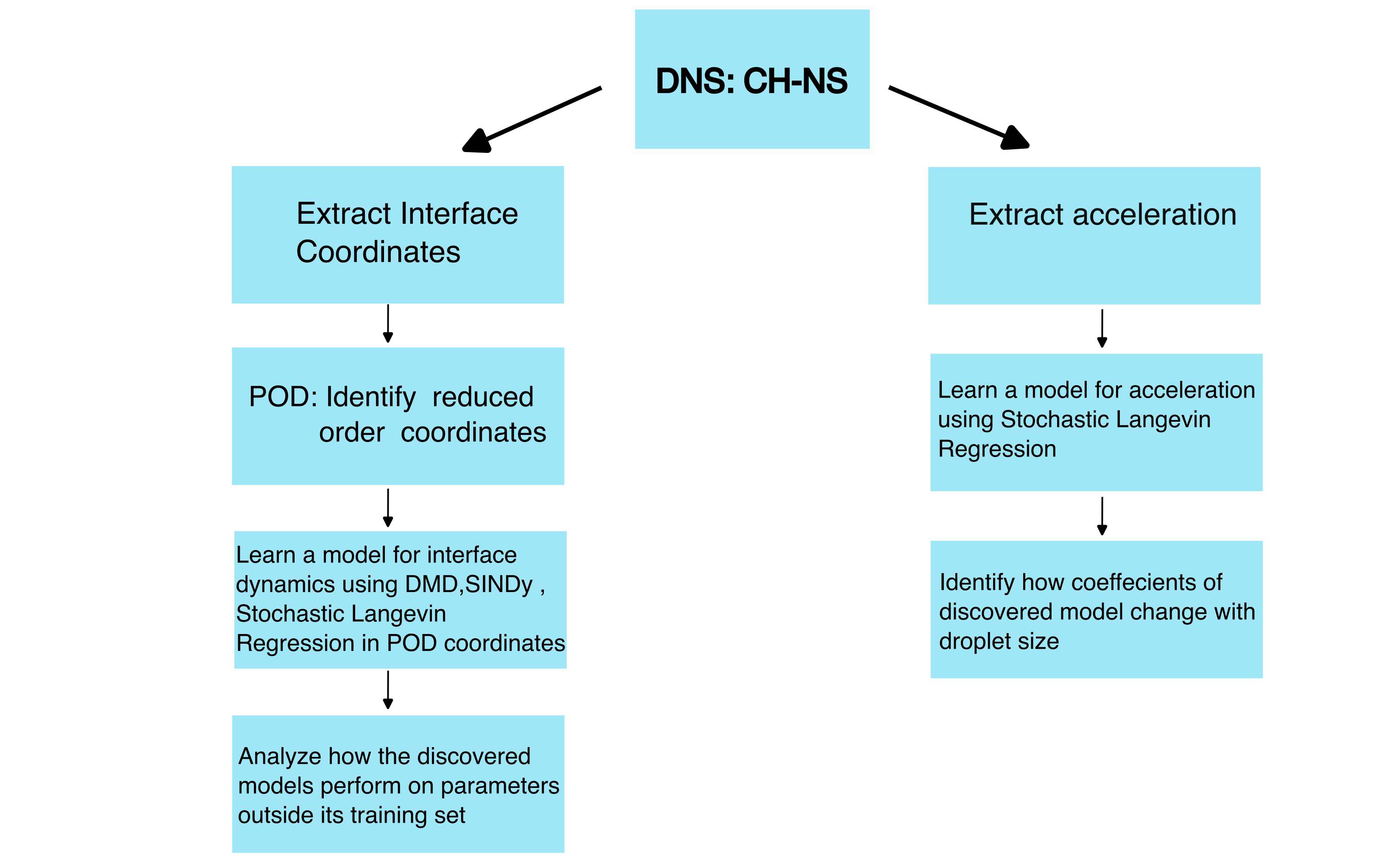}
\caption{\textbf{Description of workflow}}
\label{workflow}
\end{figure}

\paragraph{Proper Orthogonal Decomposition to identify reduced-order description:} Data-driven modeling can be split into three steps (see Figure~\ref{workflow}). In the first step, we identify the reduced-order basis using Proper Orthogonal Decomposition (POD). POD is a technique that enables us to decompose time-series data into spatial and temporal components. It can be achieved through various mathematical techniques, such as Singular Value Decomposition (SVD), the snapshot method, and convolutional autoencoders, for example. Here, we use SVD to perform POD on the DNS data to identify the dominant modes present in this system. 

\paragraph{Effective model discovery for droplet fluctuation:} The dominant modes identified through POD are used to write down a reduced-order description of the CHNS equations by truncating the modes using an appropriate optimization criterion. The truncated modes are used to develop an effective predictive model using two related approaches. In Dynamic Mode Decomposition (DMD), we assume that the modes are coupled through linear models. In contrast, in Sparse Identification of Nonlinear Dynamics (SINDy), modes can be coupled through nonlinear terms. We used DMD and SINDy to identify the equations governing droplet-perimeter fluctuations in the droplet's center-of-mass reference frame. We also use another data-driven technique, Stochastic Langevin Regression (SLR), to identify an effective model for droplet fluctuations. 

\paragraph{Effective model discovery for the center-of-mass motion:} To identify an effective model for the droplet dynamics, we also need an effective model for the center-of-mass motion, specifically its acceleration. Unfortunately, DMD and SINDy fail to provide a reliable inference mechanism for the center-of-mass acceleration. Hence, we use only SLR to identify the effective model for the center of mass. 
\medskip

First, we show that POD provides a suitable basis, interpretable as dominant deformation patterns, for constructing our model. We then present two versions of DMD, along with SINDy models, and end with the Stochastic Langevin Regression models. For each method, we first attempt to predict the interface dynamics of a droplet with higher surface tension, as droplet fluctuations are less intermittent \cite{pal2016binary} and are therefore expected to be easier to predict. We start with a surface tension corresponding to $\Lambda=200$. If the dynamics of the discovered models show statistical similarity to the DNS for this simple testing set, we explore how these models can be extended to predict interface dynamics at different surface tensions. 
\label{POD_res}

\subsection{Proper Orthogonal Decomposition to identify reduced-order description}{\label{POD_text}}
\noindent In the first step of the data-driven approach, we construct the data matrix \(H(\mathbf{\theta},t)\) (Eq.~\ref{H_matrix}) from the snapshots obtained from the direct numerical simulation of the CHNS equations. Next, by applying SVD to the data matrix \(H(t)\), we decompose it into three matrices, \(\chi\), \(\Sigma\), and \(Q\), as shown in Eq.~\ref{svd1}. 
\begin{equation}
H(\mathbf{\theta},t) 
\xrightarrow[\text{SVD}]{\quad}
\chi \Sigma Q^{\dagger} =
\sum_{j} \Sigma_{jj}\chi_j(\theta) q_j^{\dagger}(t)
\label{svd1}
\end{equation}
Here, \(\chi \in \mathbb{R}^{N \times N}\) and \(Q^{\dagger} \in \mathbb{R}^{T \times T}\) are unitary matrices that contain spatial and temporal information, respectively; \(Q^{\dagger}\) is the Hermitian conjugate of \(Q\). Specifically, the columns \((\chi_j)\) of \(\chi\) represent the dominant spatial deformation modes, and the rows $q_j^{\dagger}$ of the matrix \(Q^{\dagger}\) are the temporal coefficients describing the temporal evolution of the $j^{th}$ spatial mode, \(\chi_j\). The diagonal matrix \(\Sigma \in \mathbb{R}^{N \times T}\) contains the singular values \(\Sigma_{jj}\), which are non-negative and arranged in decreasing order. \(\Sigma_{jj}\) (Figure~\ref{fig:POD_result}(A)) quantifies the relative contribution of the product \(\chi_j(\mathbf{\theta})q_j^{\dagger}(t)\) in describing the spatiotemporal structure of \(H(\mathbf{\theta},t)\) \cite{PhysRevE.104.015206}. The singular values $\Sigma_{jj}$ can be used to identify an optimal reduced-order model. As shown in the inset of Figure~\ref{fig:POD_result}(A), the mean squared error (MSE) between the original height field and the height field reconstructed using $r$ POD modes decreases significantly up to around $r=10$ modes for a system with interfacial surface tension corresponding to $\Lambda=200$, after which further reduction in error becomes negligible. Hence, truncating the sum in Eq.~\ref{svd1} after the first ten terms provides a parsimonious reduced description of the dynamics that captures the dominant spatiotemporal features. Within this reduced basis, the droplet dynamics can be formulated by mapping from the spatial coordinate system $\mathbf{\theta} = \{\theta_i,i=1\ldots N\}$ to the modal coordinate system defined by the dominant spatial deformation modes $\{\chi_j(\mathbf{\theta}), j=1\ldots 10\}$. For example, the height field $h(\theta_i,t_k)$ at the location $\theta_i$ and time point $t_k$, which is the element $H_{ik}$ of the $H$ matrix, can then be reconstructed as:
\begin{equation}
h(\theta_i, t_k) \approx
\sum_{j=1}^{r}\Sigma_{jj} \chi_j(\theta_i) \, q_j^{\dagger}(t_k)
\label{height_reconstruct1}
\end{equation}
For a system with interfacial surface tension corresponding to $\Lambda=200$, where $r=10$. For illustration, we have shown the first three spatial modes, $\chi_{1,2,3}$ as a function of the spatial coordinates, $\theta_i$ in Figure~\ref{fig:POD_result}(B). The corresponding temporal coefficients, $q^{\dagger}_{1,2,3}$ as a function of the time points $t_k$ are shown in Figure~\ref{fig:POD_result}(C). In the following sections, we use this reduced-order description to infer the governing dynamical equations. 

\begin{figure}[h!]
    \centering
    \includegraphics[scale=0.12]{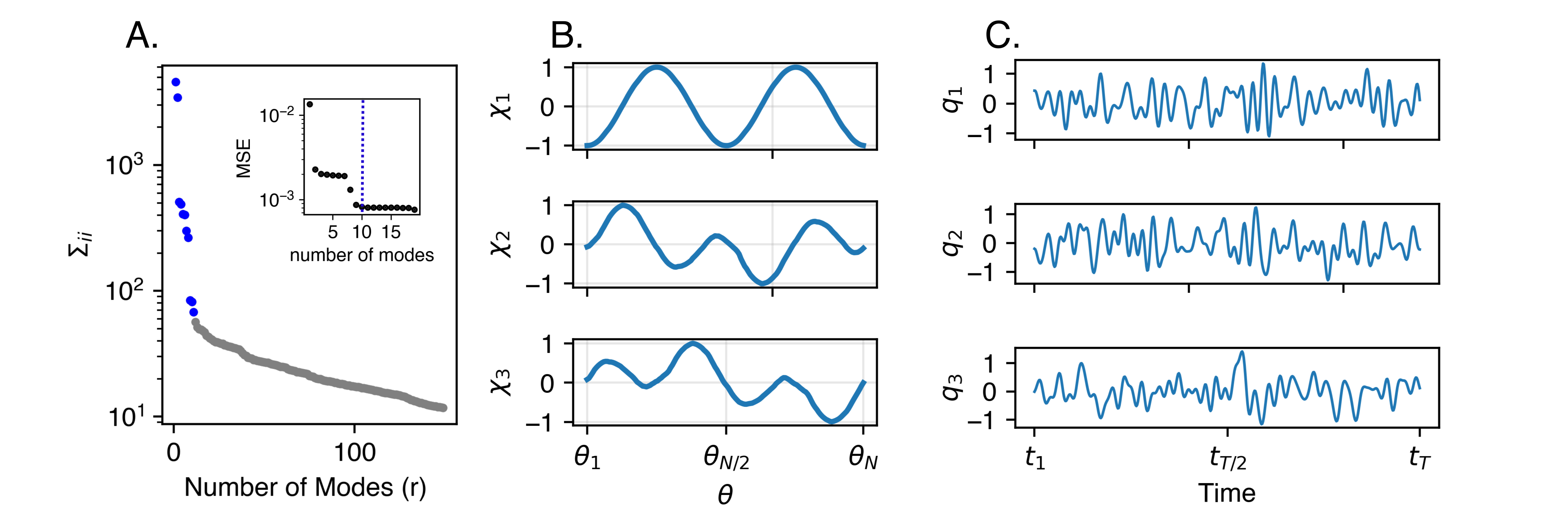}
    
    \caption{\textbf{Schematic representation of the POD using SVD:} (A, B, C) represent the components of the $\chi$, $\Sigma$, and $Q$ matrices in equation~\ref{svd1}; 
    (A) Singular values from $\Sigma_{rr}$ represent the degree of contribution of each mode to the dynamics; inset shows the mean squared error = $\frac{1}{T}\sum\limits_{i=1}^{T}|(h_i(\text{original}) - h_i(\text{truncated}))|$ when reconstructing the original $H$ matrix using mode truncation vs the number of modes used for reconstruction; 
    (B) Shows the first three highest-energy spatial modes $\chi_r$; 
    (C) Temporal modes $q_r(t)$ for the first three highest-energy modes.}
    \label{fig:POD_result}
\end{figure}

\subsection{Identifying an effective model using dynamic mode decomposition}
\label{DMD_res}

Dynamic mode decomposition (DMD) is a data-driven technique for capturing the dominant behavior of complex dynamical systems from measured data~\cite{SCHMID_2010,PhysRevE.104.015206}. DMD assumes that the temporal evolution of the system can be approximated by a linear mapping between consecutive snapshots. We define two data matrices $H_1$ and $H_2$ as
\begin{equation}
H_1 = \begin{bmatrix} h(t_1) & h(t_2) & \cdots & h(t_T) \end{bmatrix}, \quad
H_2 = \begin{bmatrix} h(t_2) & h(t_3) & \cdots & h(t_T) & h(t_{T+1}) \end{bmatrix},
\label{H1H2}
\end{equation}
and assume the existence of a linear operator $A \in \mathbb{R}^{N \times N}$ such that
\begin{equation}
H_2 = A H_1 + \epsilon,
\label{A_operator}
\end{equation}
where $\epsilon$ represents the residuals and $N$ is the state dimension. The matrix $A$ is high-dimensional and computationally expensive to handle directly. Hence, we construct its low-rank approximation $\tilde{A}$ using the singular value decomposition (SVD) of $H_1$: $H_1 \approx \tilde{\chi} \tilde{\Sigma} \tilde{Q}^\dagger$, where $\tilde{\chi} \in \mathbb{R}^{N \times r}$, $\tilde{\Sigma} \in \mathbb{R}^{r \times r}$, and $\tilde{Q}^\dagger \in \mathbb{R}^{r \times (T-1)}$ . The leading left singular vectors in $\tilde{\chi}$ form a semi-unitary basis ($\tilde{\chi}^T \tilde{\chi} = \mathrm{I}$, but $\tilde{\chi} \tilde{\chi}^T \neq \mathrm{I}$; see Appendix~\ref{SVD_s_unitary}). This yields
\begin{eqnarray}
H_1 &\approx& \tilde{\chi} \tilde{\Sigma} \tilde{Q}^\dagger, \nonumber \\
H_2 &=& A \tilde{\chi} \tilde{\Sigma} \tilde{Q}^\dagger, \nonumber \\
A &=& H_2 \tilde{Q} \tilde{\Sigma}^{-1} \tilde{\chi}^\dagger.
\end{eqnarray}

\noindent Projecting onto the reduced basis gives the low-rank companion matrix:
\begin{equation}
\tilde{A} = \tilde{\chi}^\dagger A \tilde{\chi} = \tilde{\chi}^\dagger H_2 \tilde{Q} \tilde{\Sigma}^{-1}.
\label{A_tilde}
\end{equation}

\noindent The eigenvalues $\mu_j$ of $\tilde{A}$ encode the temporal growth (or decay) rates of the modes, with corresponding eigenvectors $w_j$ as the spatial modes. The continuous-time growth rates $u_j$ are
\begin{equation}
u_j = \frac{\ln \mu_j}{\Delta t},
\end{equation}
where $\Delta t$ is the sampling interval~\cite{SCHMID_2010}. The field is reconstructed as
\begin{equation}
h(\theta_i, t_k) \approx \sum_{j=1}^r b_j w_j(\theta_i) e^{u_j t_k},
\label{DMD_reconstruct1}
\end{equation}
where the coefficients $b_j$ are obtained from the initial condition via projection onto the modes, and we truncate at a fixed low rank $r=10$ here.

\subsection{Identifying effective model using Sparse Identification of Nonlinear Dynamics}\label{SINDy}


{Droplet dynamics in turbulent flow is a highly nonlinear problem, and the failure of DMD, which employs linear predictive models, bears testimony to the underlying nonlinearity of the problem. Hence, we hypothesized that a model that would include nonlinear coupling between the spatial modes may provide better predictive capabilities. To achieve this, we employ the Sparse Identification of Nonlinear Dynamics (SINDy) framework~\cite{doi:10.1073/pnas.1517384113,Kaptanoglu2022PySINDy}, which has proven effective in discovering nonlinear evolution equations directly from time-series data~\cite{PhysRevE.104.015206,Fukami_Murata_Zhang_Fukagata_2021,PhysRevLett.129.258001,doi:10.1073/pnas.2016708118}.}

\subsubsection{Set up of the SINDy model}
{ To learn an effective nonlinear model using SINDy, we work in the reduced basis developed in Section~\ref{POD_text}. Akin to DMD, in the reduced framework, the $H$-matrix is written as: 
\begin{equation}
    H \approx \tilde{\chi}\tilde{\Sigma}\tilde{Q}^{\dagger}.
    \label{sindy_XSQ}
\end{equation}
However, in contrast to DMD, $\tilde{Q}^{\dagger}\in \mathbb{R}^{T\times r}$. The nonlinear coupling between the modes is achieved by assuming the following evolution equation of the temporal coefficients $\tilde{Q}^{\dagger}$:
\begin{equation}
    \dot{\tilde{Q}}^{\dagger} = \Theta(\tilde{Q}^{\dagger})\Xi \label{SINDy_starting},
\end{equation}
where $\Theta(\tilde{Q}^{\dagger}) \in \mathbb{C}^{T\times m}$, described below, is a library of nonlinear couplings between the temporal coefficients, and $\Xi \in \mathbb{R}^{m\times r}$ is the coefficient matrix. Here $m$ is the number of terms in the library. The goal of SINDy is to infer the coefficient matrix $\Xi$ from the time series data.}

{To find $\dot{\tilde{Q}}^{\dagger}$, we use a central difference scheme. We define the matrix $\Theta \in \mathbb{R}^{T \times m}$ as a library of candidate nonlinear terms to be included in the discovered dynamical equations, where $m$ denotes the total number of such terms. For example, for $r = 10$ modes, to construct a library with polynomial terms up to quadratic order in $q_j$, we need one constant term, ten linear terms, and fifty-five quadratic terms, i.e., $m = 66$. Using higher-order terms greatly increases the size of the library and reduces the interpretability of the discovered model. The precise arrangement of the elements in $\Theta$ is detailed in Eq.~\ref{Theta_elements1_modified}.

\begin{equation}
\Theta(\tilde{Q}) =
\begin{bmatrix}
1 & q_{11}  & \dots & q_{1,10} & q_{11}^2  & \dots & q_{1,10}^2 & q_{1,1} q_{1,2}& \dots & q_{1,9}q_{1,10} \\
1 & q_{2,1} & \dots & q_{2,10} & q_{2,1}^2& \dots & q_{2,10}^2 & q_{2,1} q_{2,2}& \dots & q_{2,9}q_{2,10} \\
\vdots & \vdots & \ddots & \vdots & \vdots & \ddots & \vdots & \vdots & \ddots & \vdots \\
1 & q_{T,1}  & \dots & q_{T,10} & q_{T,1}^2 & \dots & q_{T,10}^2 & q_{T,1} q_{T,2}& \dots & q_{T,9}q_{T,10}
\end{bmatrix}
\label{Theta_elements1_modified}
\end{equation}

Using this library, we aim to infer a $\Xi$ matrix that satisfies Eq.~\ref{SINDy_starting} and maximizes the number of zero elements. This type of regression is termed sparse regression. The sparsity is enforced here using Stepwise Sparse Regression (SSR)~\cite{10.1063/1.5018409}. To do so, we follow the following algorithm:

\begin{itemize}
\item First, we perform a non-sparse regression to find a matrix $\Xi_1$ that minimizes the least square error: $\|  \tilde{\dot{Q}}^\dagger - \Theta \Xi_1 \|_{2}
$, where the subscript 2 implies the $L_2$ norm.

\item Next, the lowest term of $\Xi_1$ is set to zero, which generates the second iteration of the matrix, termed $\Xi_2$. 

\item Next, we perform the non-sparse regression on $\Xi_2$.

\item At each step, we calculate the predictive performance of the inferred $\Xi$ matrix using the KL divergence on a cross-validation set. 
\item The sparsest solution that does not increase the cross-validation error by threefold is chosen \cite{10.1063/1.5018409}.

\end{itemize}

This formulation enables the direct identification of interpretable nonlinear governing equations from time series data, potentially facilitating the effective modeling of complex systems. }

\subsection{Identifying the effective model using Stochastic Langevin Regression}
\label{SLRinterface}

Although SINDy models perform better than DMD models at predicting the deformation statistics of the droplet interface, we find that a large number of coupled equations is still required to achieve consistent predictive power. In addition, inconsistent performance in predicting interface dynamics across different surface tensions indicates a lack of interpretability in the models. This likely arises because, in turbulent flows, droplet deformations exhibit irregular fluctuations that are characteristic of intermittency \cite{pal2016binary}. Such behavior makes purely deterministic modeling infeasible, particularly within reduced-order representations, where truncated modes may encode these intermittent bursts \cite{NOACK_AFANASIEV_MORZYŃSKI_TADMOR_THIELE_2003, Schmidt_Schmid_2019}.

In this section, we explore a stochastic modeling approach based on Stochastic Langevin Regression (SLR)~\cite{Landau1987Fluid,fluids5030108}, which has been successfully applied to low-dimensional representations of turbulent systems \cite{doi:10.1098/rspa.2021.0092,WANG2022244}. As in the previous Section~\ref{SINDy}, we work in the reduced coordinate system obtained from POD and seek to predict the dynamics of the temporal coefficients $q_j$ obtained from the POD analysis. Unlike SINDy, we do not assume that the trajectories are governed by coupled deterministic equations. Instead, each temporal coefficient $q_j$ is assumed to evolve as a stochastic process characterized by a drift term and a generally nonlinear stochastic forcing term. Accordingly, the dynamics of $q_j$ are modeled as
\begin{equation}
dq_j = f(q_j)\,dt + \sigma(q_j)\,dW(t),
\label{stoch_langevin}
\end{equation}
where $f(q_j)$ is the drift term, $\sigma(q_j)$ is a state-dependent diffusion coefficient that may include nonlinear dependencies, and $W(t)$ is a Wiener process (delta-correlated white noise).

\noindent To estimate the drift and diffusion components, the distribution of the stochastic trajectory of $q_j$ is constructed using a histogram, whose optimal bin width is determined from a Gaussian kernel density estimate (KDE). Figures~\ref{SLR_demo}(A,B) illustrate this for a simple one-dimensional trajectory of a particle in a double-well potential. The trajectory is then sampled at a lag time $\tau$, determined following the Kramers--Moyal averaging procedure (Appendix~\ref{SLR_details} and Refs.~\cite{PhysRevE.83.066701,doi:10.1098/rspa.2021.0092,FRIEDRICH201187}). This timescale separation supports approximate Markovianity and enables the drift and diffusion coefficients to be estimated from the sampled data using the Kramers--Moyal expansion (Appendix~\ref{SLR_details}):

\begin{equation}
f_{\mathrm{KM}}(q) \approx \frac{1}{\tau} \left\langle q_{t+\tau} - q_t \,\big|\, q_t = q_i \right\rangle,
\label{fkm}
\end{equation}
\begin{equation}
\sigma_{\mathrm{KM}}(q) \approx \frac{1}{2\tau} \left\langle \left(q_{t+\tau} - q_t\right)^2 \,\big|\, q_t = q_i \right\rangle.
\label{akm}
\end{equation}

Here $\langle \cdot \mid q_t = q_i \rangle$ denotes an average over all observed increments whose initial value $q_t$ lies in a bin centered around $q_i$. In Figure~\ref{SLR_demo}(C), $f(q_j)$ and $\sigma(q_j)$ are calculated for the given trajectory and shown as scatter plots. In the stochastic Langevin regression approach \cite{doi:10.1098/rspa.2021.0092,10.1063/1.5018409}, the functional forms of $f(q_j)$ and $\sigma(q_j)$ are modeled using SINDy (Figure~\ref{SLR_demo}(C), orange and blue lines). 

This regression is carried out by defining a cost function that balances the drift and diffusion errors with a Kullback--Leibler divergence penalty that enforces consistency between the model distribution and the data distribution, defined as:
\begin{equation}
\text{Cost} = 
\underbrace{\sum_{i} w_f |f_\tau^i - f_{\mathrm{KM}}^i|^2}_{\text{Drift error}} +
\underbrace{\sum_{i} w_a |\sigma_\tau^i - \sigma_{\mathrm{KM}}^i|^2}_{\text{Diffusion error}} +
\underbrace{\lambda_{KL} D_{KL}(p_{\text{data}} \| p_{\text{model}})}_{\text{Distribution error}}.
\label{cost_SLR}
\end{equation}

\begin{figure}[h!]
    \centering
    \includegraphics[scale=0.16]{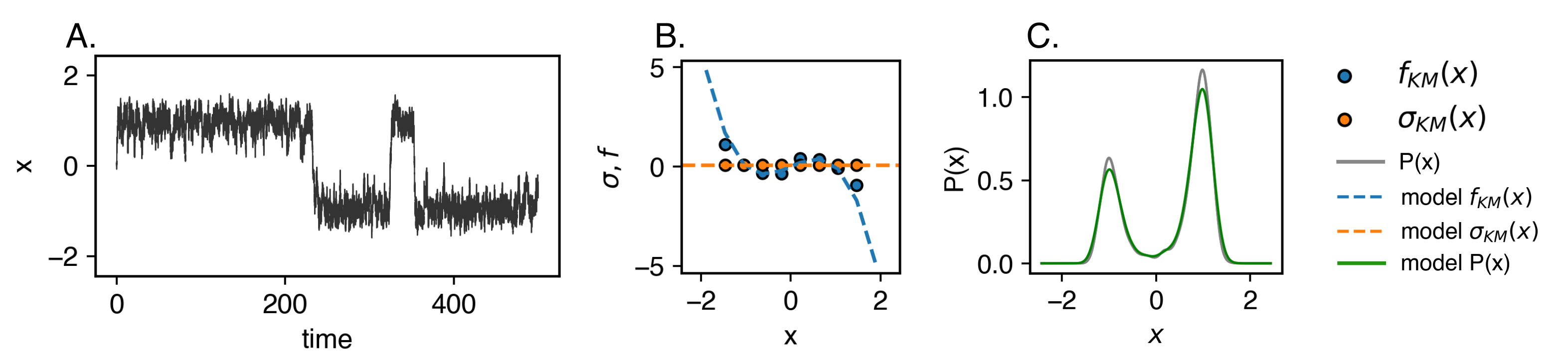}
    \caption{\textbf{Simple example of the implementation of SLR.}
(A) A representative stochastic trajectory of a one-dimensional particle evolving in a double-well potential.
(B) Scatter plots illustrating the estimation of the drift term $f_{\mathrm{KM}}$ and diffusion coefficient $\sigma_{\mathrm{KM}}$ from the stochastic trajectory $x$ within each bin, computed using Eqs.~\eqref{fkm} and~\eqref{akm}. Dashed lines denote the corresponding model inferred via SINDy.
(C) Comparison of the trajectory predicted by the inferred model with the testing dataset to assess statistical similarity; the model minimizing the cost function defined in Eq.~\eqref{cost_SLR} is selected.
}
    \label{SLR_demo}
\end{figure}

The weight applied to the drift error is $w_f$, and the estimated drift at histogram bin $i$ is $f_{\tau}^i$. The empirical drift estimate at bin $i$, obtained via the Kramers--Moyal method, is $f_{\mathrm{KM}}^i$. Similarly, the diffusion error is weighted by $w_a$, with the model estimate $\sigma_{\tau}^i$, and the empirical diffusion estimate $\sigma_{\mathrm{KM}}^i$.

The weights $w_a$, $w_f$, and $\lambda_{KL}$ are implemented using inverse-variance weighting based on the standard error of the Kramers--Moyal coefficients \cite{doi:10.1098/rspa.2021.0092}. This ensures that well-sampled regions of state space (where drift and diffusion estimates are statistically reliable) have greater influence on the regression, while poorly sampled or noisy regions are automatically down-weighted.

In this SSR procedure, the sparsest model is obtained by initializing the library of terms for the drift and stochastic components using polynomial terms up to fifth order,
\[
[q_j, q_j^2, \ldots, q_j^5].
\]
The SSR procedure employed in SLR differs from that used to model deterministic equations, as the smallest coefficient in a stochastic model is not necessarily the least significant \cite{doi:10.1098/rspa.2021.0092}. For example, given a stochastic equation:
\begin{equation}
\dot{x} = x + 0.001x^3 + \text{noise},
\end{equation}
eliminating the cubic term yields a model with a single stable equilibrium at $x=0$, a Gaussian stationary probability density function, and no mechanism for bistability or heavy-tailed fluctuations. As a result, the reduced model exhibits substantial discrepancies with the empirical statistics, including the stationary PDF, the finite-time Kramers--Moyal coefficients, and the distribution of residence times. In contrast, for a purely deterministic system, neglecting the cubic term—despite its nonlinearity—typically constitutes only a weak perturbation of the dynamics, leaving the local stability structure largely unchanged.

For this reason, terms are iteratively dropped by removing the term that produces the smallest increase in the cost function at each step. The cost function is recalculated after each iteration, and the equation corresponding to the minimum cost is selected \cite{doi:10.1098/rspa.2021.0092}.

\section{\label{sec:level4}RESULTS}
In this section, we present our results showing the performance of the data-driven models in use.

\subsubsection{Effective DMD model for droplet interface dynamics in turbulent flow}

The verification of our implementation of the DMD algorithm is provided in Appendix~\ref{DMD_verify}.
Dynamic Mode Decomposition (DMD) was applied to the height matrix \(H\) (Eq.~\eqref{H_matrix}) generated from the CHNS equations for a surface tension \(\Lambda = 200\).
The discrete-time eigenvalues \(\mu_j\) of the reduced operator \(\tilde{A}\) (see Eq.~\eqref{A_tilde}) are shown in Figure~\ref{fig:DMD_result1}(A) (red dots).
We observe that the magnitudes of the eigenvalues satisfy \(|\mu_j| < 1\), while their imaginary parts are small but nonzero.

To interpret these eigenvalues in terms of the underlying continuous-time dynamics, each eigenvalue is written in polar form,
\begin{equation}
\mu_j = |\mu_j| e^{i\theta_j}.
\end{equation}
The corresponding continuous-time growth rates \(u_j\) are obtained via the standard DMD mapping
\begin{equation}
u_j = \frac{\ln(\mu_j)}{\Delta t}
    = \frac{\ln|\mu_j|}{\Delta t} + i\frac{\theta_j}{\Delta t},
\end{equation}
where \(\Delta t\) denotes the sampling interval.
The real part \(\text{Re}(u_j)=\ln|\mu_j|/\Delta t\) governs the exponential growth or decay of each mode, while the imaginary part \(\text{Im}(u_j)=\theta_j/\Delta t\) corresponds to its oscillation frequency~\cite{schmid2022dynamic}.
Since \(|\mu_j|<1\) for all modes, the real parts of \(u_j\) are strictly negative, indicating that all DMD modes decay rapidly in time.
Hence, the deformation parameter computed from the height field reconstructed from the calculated modes (Eq.~\ref{DMD_reconstruct1}) dies very rapidly. In contrast, the deformation parameter calculated from DNS shows large fluctuations (Figure~\ref{fig:DMD_result1}(B)).

To assess the improvement in predictive capability, we repeated the analysis using the Hankel DMD framework (see Appendix~\ref{HankelDMD}). In Figure~\ref{fig:DMD_result1}(A) (blue dots), the discrete-time eigenvalues computed via Hankel DMD are distributed approximately along the unit circle in the complex plane, exhibiting predominantly imaginary values consistent with fast, oscillatory dynamics. Reconstruction of the height field shows that the droplet deformation parameter now oscillates, in qualitative agreement with the DNS results. However, the amplitude of the predicted deformation parameter remains significantly underestimated relative to DNS, as shown in Figure~\ref{fig:DMD_result1}(B) (blue line). This result shows that DMD-based models cannot capture droplet deformation dynamics.

\begin{figure}[h!]
    \centering
    \includegraphics[scale=0.12]{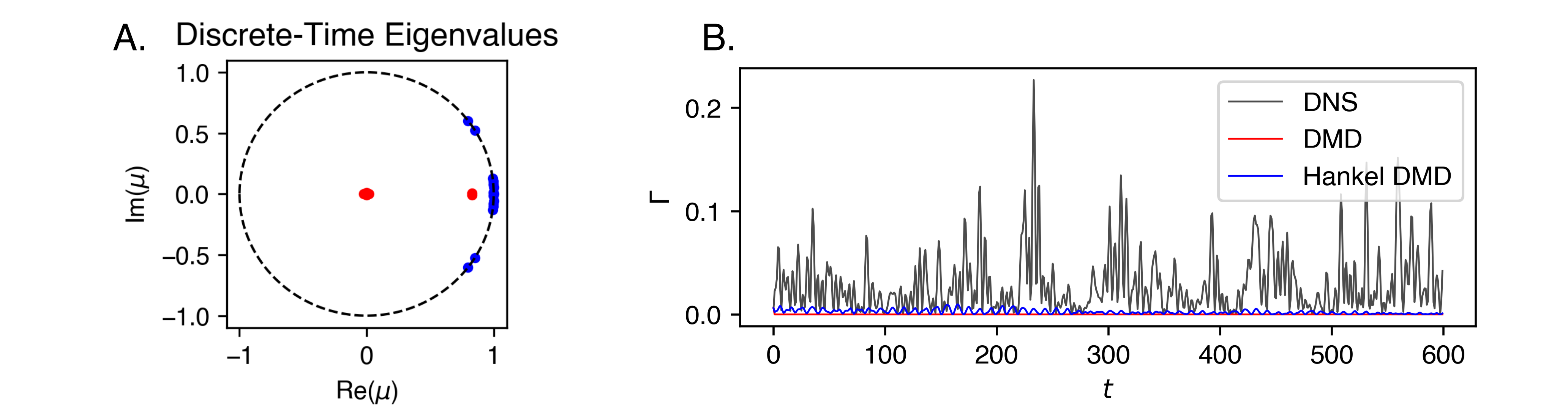}
    \caption{\textbf{DMD for interface fluctuations:}
    (A) Discrete-Time Eigenvalues of DMD evolution matrix for DMD (red dots) and Hankel DMD (blue dots). $\mu=1$ (point on a circle) signifies purely oscillatory modes, $\mu>1$ signifies growing modes, and  $\mu<1$ signifies decaying modes \cite{schmid2022dynamic}. (B) Prediction of deformation parameter obtained from DMD models (red) and Hankel DMD (blue) compared to DNS (black) for the test dataset.}   \label{fig:DMD_result1}
\end{figure}

\subsubsection{Effective SINDy model for droplet fluctuation in turbulent flow}

The inability of the DMD model to accurately capture the dynamics of the droplet in turbulence motivated us to apply the SINDy model to study the dynamics of the droplet. The verification of the code used for the SINDy algorithm is provided in the Appendix~\ref{verify_SINDY}. We apply SINDy in the reduced coordinate system to learn the dynamics of the temporal coefficients $q_j$, which are rows of the $\tilde{Q}$ matrix calculated from the POD of the matrix $H$ following Eq.~\eqref{svd1}. In Figure~\ref{fig:sindy_results}(A), we show the learned coefficient matrix $\Xi$ as a color-coded array. We see that the dynamics of $q_j$ depend mostly linearly on each other in a coupled manner, with a few quadratic terms with lower coefficients also contributing to the dynamics. We can now solve the discovered coupled ODEs using the LSODA (Livermore Solver for Ordinary Differential Equations with Automatic method switching) algorithm \cite{petzold1983automatic}, the same initial conditions as the testing set, and obtain the evolution of the $q_j$ temporal coefficients over time. To relate these results to the measurable quantity of the droplet deformation parameter $\Gamma$, we store these predicted temporal coefficients in a matrix $\tilde Q_{\text{model}}$ and reconstruct the height field from these predicted temporal coefficients using the equation

\begin{equation}
    H = \tilde{\chi} \tilde{\Sigma} \tilde{Q}_{\text{model}}^{\dagger}
    \label{H_reconstruct}
\end{equation}

In Figure~\ref{fig:sindy_results}(B), the droplet deformation parameter is calculated from the reconstructed height field and is compared with the original droplet deformation data for the testing set. Figure~\ref{fig:sindy_results}(C) compares the distribution of $\Gamma$ to check the statistical similarity of the SINDy-predicted deformation parameters. The SINDy framework facilitates the identification of reduced-order models that span a spectrum of complexity—from simple representations with fewer modes and higher errors to more elaborate, computationally intensive models that risk overfitting~\cite{MEHTA20191}. To assess this trade-off, we compute the Kullback–Leibler divergence \cite{shlens2014noteskullbackleiblerdivergencelikelihood}($D_{\mathrm{KL}}$) of the distribution of the predicted droplet deformation parameter ($\Gamma$) compared to the testing set:

\begin{equation}
D_{\mathrm{KL}}\left(\Gamma_{\text{SINDy}} \parallel \Gamma_{\text{DNS}}\right)
= \sum_{x} \Gamma_{\text{SINDy}}(x) \, \log\!\left(\frac{\Gamma_{\text{SINDy}}(x)}{\Gamma_{\text{DNS}}(x)}\right)
\label{KLD_eq}
\end{equation}

Figure~\ref{fig:sindy_results}(D) presents the variation of the KL divergence as a function of the number of modes retained in the model. The minimum KL divergence is achieved for a model including the ten most energetic modes ($r=10$), consistent with the trend observed in the POD reconstruction error discussed in Sec.~\ref{POD_text}. Despite this optimal choice, the predicted temporal evolution of $\Gamma$ deviates noticeably from the measured deformation parameter, although it provides a significant improvement over the DMD-based prediction (Figure~\ref{fig:DMD_result1}(B)). Moreover, the statistical characteristics of the predicted and measured time series exhibit close similarity, as evidenced by their comparable distributions
(Figure~\ref{fig:sindy_results}(D)).
\begin{figure}[h]
    \centering
    \includegraphics[scale=0.11]{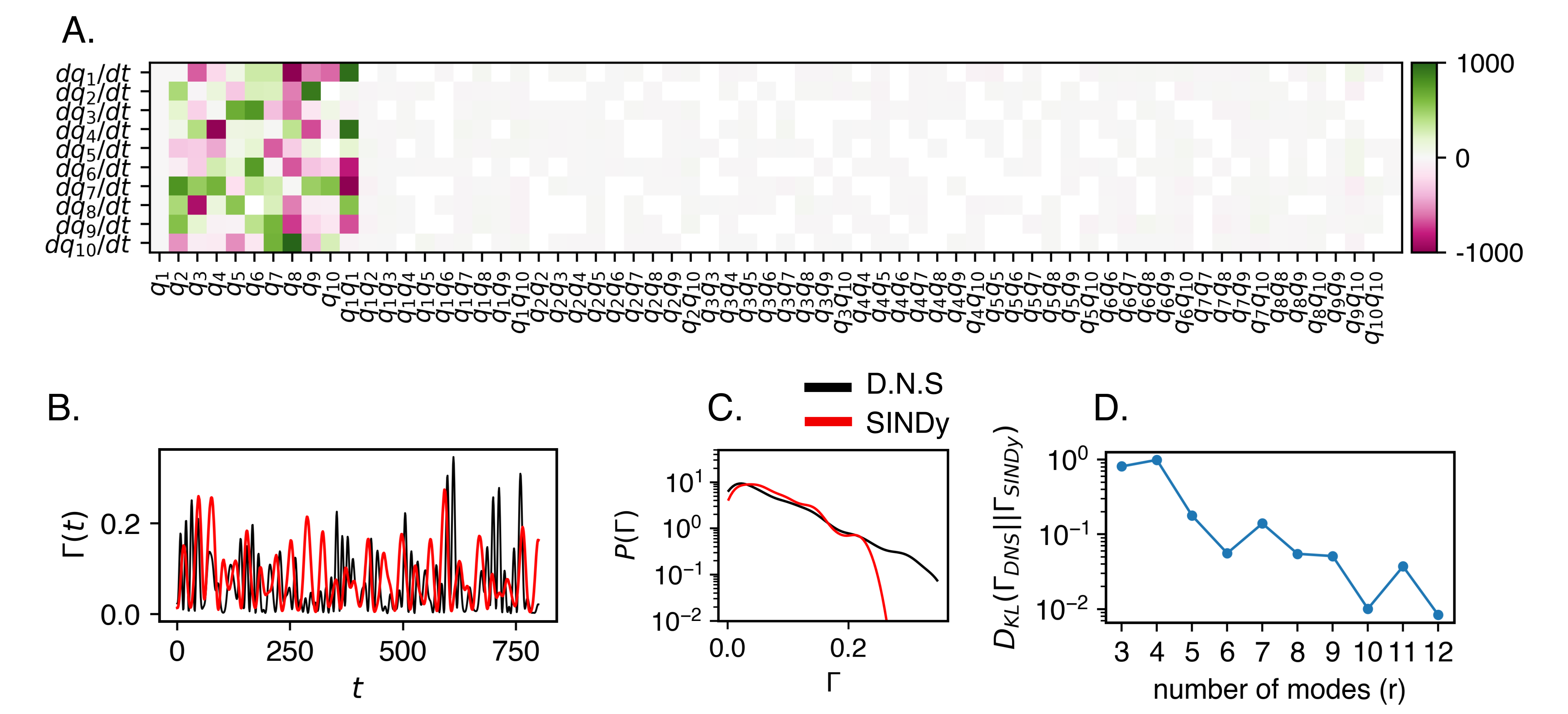}
    \caption{ \textbf{SINDy for interface fluctuations:}
(A) Coefficient matrix $\Xi$ obtained from sparse regression (color-coded); each row corresponds to a predicted time derivative $\dot{q_j}$, and each column denotes the contribution of a candidate term in the library matrix, as defined in Eq.~\ref{SINDy_starting}. (B) Comparison of the droplet deformation parameter $\Gamma$ reconstructed from the height field predicted by SINDy (see Eq.~\ref{height_reconstruct1}) with the corresponding value computed directly from the DNS. (C) Probability density function (PDF) of the deformation parameter $\Gamma$  shown in panel~(B). (D) KL Divergence (Eq.~\ref{KLD_eq}) for $\Gamma$ obtained from SINDy model constructed from $r$ modes compared with DNS testing data}
   \label{fig:sindy_results}
\end{figure}

\subsubsection{SINDy does not generalize well to other surface tension}\label{Generalisation}
{
For the discovered model to be practically useful, it must generalize across physical parameters. For example, a model trained at $\Lambda = 200$ should accurately predict droplet dynamics at other surface tension values. Although no analytical relation is available for the dependence of the $\Xi$-matrix coefficients on $\Lambda$, trends can be inferred directly from data by examining how the spatial, temporal, or energetic components of the SVD  vary with surface tension \cite{Sato_Schmidt_2025}.

Direct numerical simulations (DNS) of the CHNS equations were performed for different values of $\Lambda$, and POD was applied to extract the mode amplitudes $q_j$, spatial structures $\chi_j$, and singular values $\Sigma_{jj}$, as shown in Figures~\ref{fig:sindy_not_general}(A--C). No systematic variation of $q_j$ or $\chi_j$ with $\Lambda$ is observed; however, $\Sigma_{jj}$ shows a clear dependence on surface tension. The singular values, which represent the energy content of each mode, increase as surface tension decreases. Motivated by this observation and methods proposed in related works \cite{10.1098/rsos.240995,10.1063/5.0266302}, the discovered SINDy models can be extrapolated to predict the interface dynamics of droplets with different surface tensions as follows:

\begin{itemize}
\item The associated spatial basis $\chi_{\text{train}}$ is obtained from the POD decomposition of the H matrix for the surface tension the model is trained on (Eq.~\ref{sindy_XSQ}).
\item The SINDy model provides a complete description of the temporal coefficient matrix $\tilde Q$ for the training surface tension $\Lambda_{\text{train}}$, denoted $\tilde Q_{\text{train}}$ (Eq.~\ref{SINDy_starting}).

\item To test a model trained at a given surface tension on a different surface tension, the height field is reconstructed using the discovered model's temporal coefficients $Q_{\text{train}}$, the basis $\chi_{\text{train}}$ from the training set, and the singular values $\Sigma_{\Lambda}$ corresponding to the target surface tension:
\begin{equation}
H_{\Lambda} \approx \tilde{\chi}_{\text{train}} \tilde{\Sigma}_{\Lambda} \tilde{Q}^{\dagger}_{\text{train}}
\label{sindy_XSQ}
\end{equation}
\item From the reconstructed height fields $H_{\Lambda}$, the droplet deformation parameter $\Gamma$ is computed, and its distribution is compared with the testing set obtained from DNS. The discrepancy between the predicted and DNS distributions is quantified using the Kullback–Leibler (KL) divergence (Figure~\ref{fig:sindy_not_general}(D))\cite{shlens2014noteskullbackleiblerdivergencelikelihood}. 
\end{itemize}

 \begin{figure}[h!]
    \centering
    \includegraphics[scale=0.13]{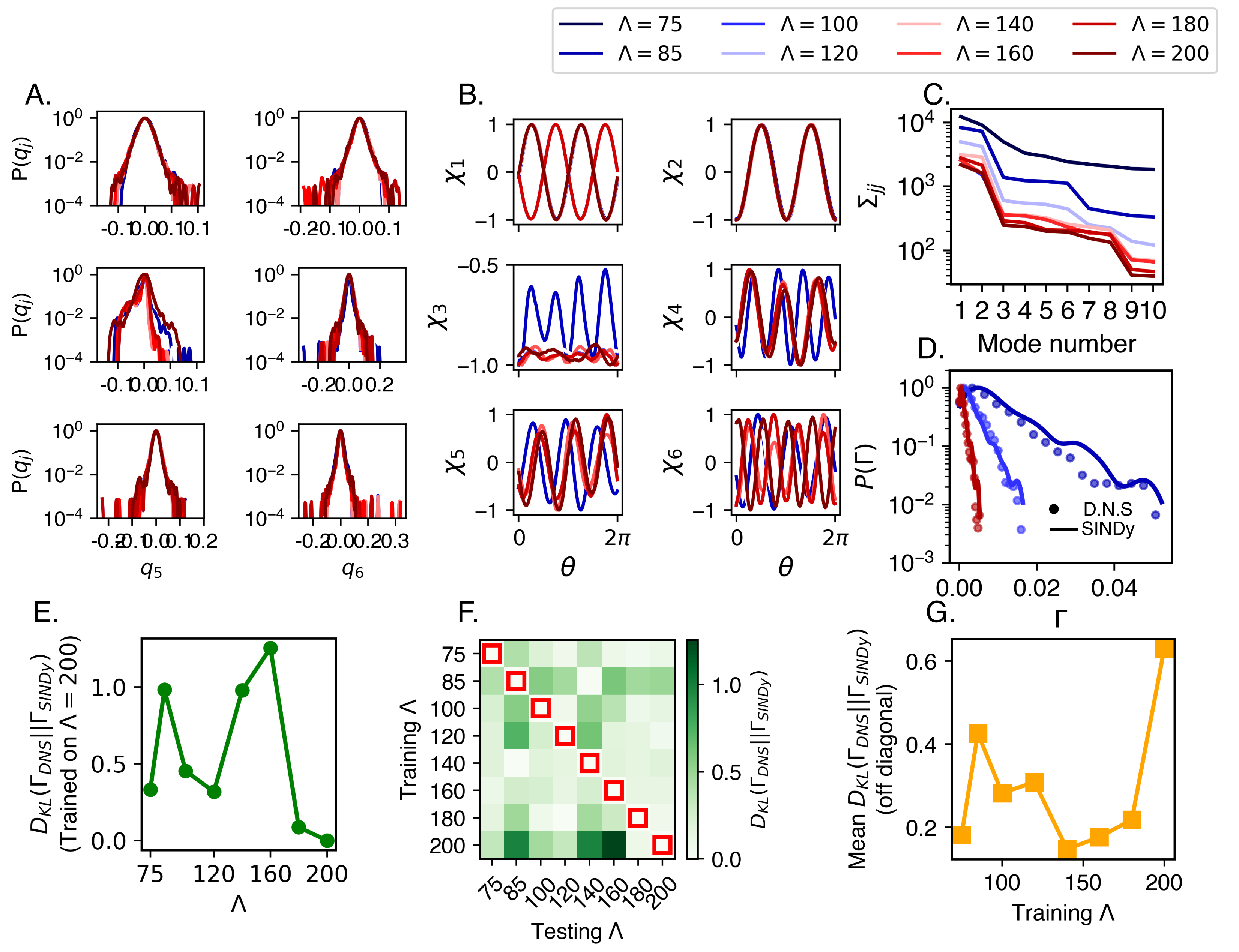}
    \caption{\textbf{Analysis of the generalisability of the discovered models:}
(A) Distribution of the temporal coefficients $q_j$ for the six highest-energy modes shows no distinct trends as surface tension is changed.
(B) Spatial deformation modes $\chi_j$ of the highest-energy modes show no distinct trends as surface tension is changed.
(C) Diagonal elements $\Sigma_{jj}$ from POD of the deformation matrix $H$
show an increase in the energy of each mode as surface tension decreases. 
(D) Probability distributions for the deformation parameter $\Gamma$ from SINDy models (solid curves) trained at $\Lambda=200$ and generalized to $\Lambda=100 \text{ and } 85$
compared with distributions from DNS (dotted curves).
(E) Mismatch between the distributions in (D) quantified by the KL divergence.
(F) The analysis in E is performed on data trained at different surface tensions and stored in a KL divergence matrix quantifying the 
performance of models trained at surface tensions (y-axis) when predicting dynamics at other surface tensions (x-axis). Red squares mark the training datasets.  
(G) The mean generalization KL divergence for each row in (F) shows that models trained at intermediate surface tensions ($\Lambda = 140$) generalize best across conditions. 
}
\label{fig:sindy_not_general}
\end{figure}
\noindent In Figure~\ref{fig:sindy_not_general}(E), the KL divergence of a SINDy model trained at $\Lambda=200$ is plotted to evaluate its predictive performance on interface dynamics corresponding to other surface tensions. The increase in KL divergence when the model is applied to other surface tensions indicates that the trained model does not generalize well to other surface tensions. In Figure~\ref{fig:sindy_not_general}(F), we repeat this same analysis for models trained at different surface tensions ($\Lambda_{\text{train}}$) when they are used to predict interface dynamics at other surface tensions ($\Lambda_{\text{test}}$) and represent the KL divergences in a matrix.
In Figure~\ref{fig:sindy_not_general}(G), we analyze this matrix by calculating the mean KL divergence for each surface tension apart from the training set (the non-diagonal entries of each row). Overall, we conclude that SINDy models generally fail to predict dynamics that generalize consistently across different surface tensions.}

\subsubsection{Effective SLR model for droplet fluctuation in turbulent flow}{\label{SLR_text}}
The lack of consistent predictability of the SINDy model when used to study the dynamics of droplets across different surface tensions, inspired us to use a stochastic approach in implementing the Langevin Regression. A detailed verification of our SLR code is provided in Appendix~\ref{verify_SLR}. We perform sparse SLR on the temporal coefficients
$q_j$ . Figure~\ref{fig:stochastic_plot}(A) presents the cost function across successive iterations of the SSR procedure when applied to the first two temporal coefficients $q_1$ and $q_2$. Models containing six terms were selected as they provide an optimal compromise between minimizing the cost function and maintaining a compact representation.
The discovered stochastic differential equations take the form

\begin{equation}
dq_j = (c_{j1} q_j + c_{j3} q_j^3)dt + (K_{j0} + K_{j1}q_j + K_{j2}q_j^2 + K_{j3}q_j^3)dW_t,
\label{stoch_eqns_interface}
\end{equation}
 where  $c_{jk} $ and $K_{jk} $ denote the drift and diffusion coefficients associated with the $ k^{\text{th}} $-order polynomial terms of the $ j^{\text{th}}$ temporal coefficient. 
The structure of the discovered equations admits a physical interpretation. The cubic drift term $c_{j1} q_j + c_{j3} q_j^3$ is reminiscent of a  restoring force derivable from an effective quartic potential $V(q_j) = -\frac{c_{j1}}{2}q_j^2 - \frac{c_{j3}}{4}q_j^4$. The state-dependent diffusion $\sigma(q_j) = K_{j0} + K_{j1}q_j + K_{j2}q_j^2 + K_{j3}q_j^3$ implies that the effective noise experienced by each deformation mode depends on the instantaneous amplitude of that mode. This is physically reasonable: larger deformations expose the interface to stronger velocity gradients in the turbulent flow, thereby amplifying the stochastic forcing \cite{Perlekar2017}. The constant term $K_{j0}$ represents a baseline turbulent forcing independent of deformation amplitude, while the higher-order terms encode the coupling between deformation amplitude and turbulent fluctuations. Similar multiplicative noise structures have been reported in stochastic models of turbulent transport~\cite{Friedrich2011} and interface fluctuations~\cite{callaham2021nonlinear}.

 The identified stochastic equations for the temporal coefficients were integrated in time using the Euler--Maruyama scheme~\cite{maruyama1955continuous}. On comparing the distributions of the discovered models' predictions of $q_j$s with the DNS, we find that the model dynamics are statistically similar to the original simulation (Figure~\ref{fig:stochastic_plot}(B)). From these simulated temporal coefficients, the height field ($H$) was reconstructed via Eq.~\ref{H_reconstruct}, enabling computation of the droplet deformation parameter and hence $\Gamma$. Model performance was assessed by comparing the probability distribution of $ \Gamma $ obtained from the reconstructed dynamics with that of the original droplet system using the KL Divergence. Figure~\ref{fig:stochastic_plot}(C) illustrates this comparison as the number of temporal modes $r$ included in the reconstruction is increased. We observe that inclusion of more than two modes yields negligible improvement. Thus, the SLR framework effectively reduces the system dynamics to two stochastic differential equations while preserving the statistical properties of the droplet deformation.

\begin{figure}[h!]
    \centering
    \includegraphics[scale=0.11]{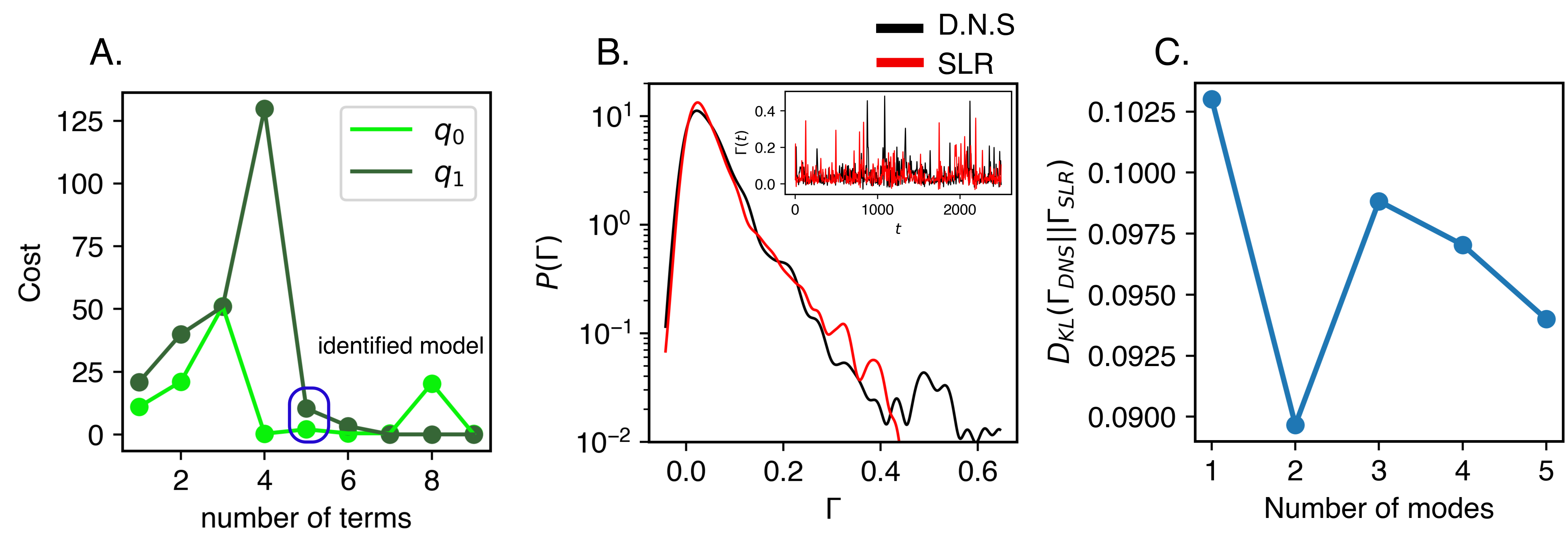}
    \caption{\textbf{Stochastic Langevin Regression for interface fluctuation:} (A) The cost function [see Eq.~\ref{cost_SLR}] at each step of sparse symbolic regression (SSR), trained on droplet data with
varying initial droplet radius $L$. During each SSR iteration, one library term is removed until a noticeable increase in the cost function
occurs. (B) The $H$ matrix can be reconstructed using Eq.~\ref{H_reconstruct}, and the distribution of the deformation parameter ($P(\Gamma)$) obtained from the SLR model is compared to that of the DNS. The inset shows the time evolution of $\Gamma$ from which $P(\Gamma)$ is obtained.  (C) Comparing the KL Divergence of  $\Gamma$ predicted from an SLR model with a different number of modes shows that a model with just two modes is enough to minimize the KL Divergence.}
    \label{fig:stochastic_plot}
\end{figure}

\subsubsection{SLR generalizes well to other surface tensions}
We repeat the same analysis done in Section~\ref{Generalisation} using models obtained from SLR. In Figure~\ref{fig:SLR_general}(A), we find that for data trained on surface tension $\Lambda=200$, the performance of the models on other surface tensions is also much better. We repeat this analysis for models trained on different surface tensions and show the results in Figure~\ref{fig:SLR_general}(B). Analyzing the mean KL divergences for the model's performance in predicting interface dynamics of droplets with a surface tension where the model was not trained on, shows that SLR models generalize to other surface tensions much better than SINDy models.

\begin{figure}[h!]
    \centering
    \includegraphics[scale=0.17]{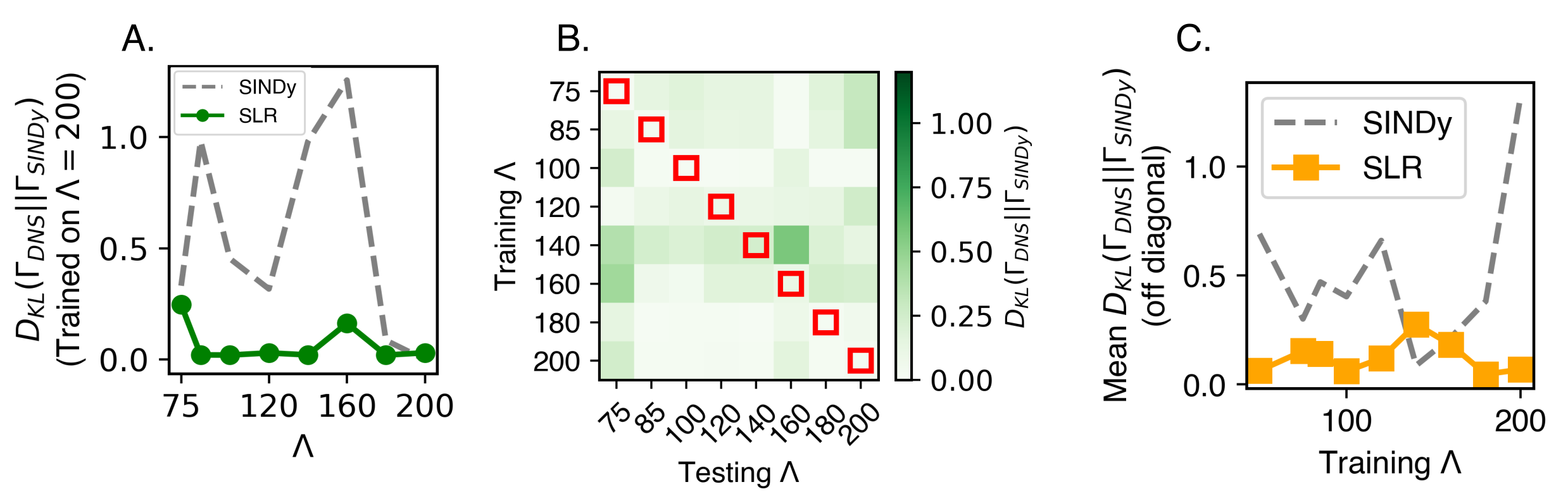}
    \caption{(A) KL divergence of a SLR model trained on $\Lambda=200$(green) when used to predict interface dynamics for other surface tensions. The same analysis for SINDy models trained in Section~\ref{SINDy} is shown in grey. (B) The analysis in A is performed on data trained at different surface tensions and stored in a KL divergence matrix quantifying the 
performance of models trained at  surface tensions (y-axis) when predicting dynamics at other surface tensions (x-
axis). Red squares mark the training datasets.
(C) The mean KL divergence of the off-diagonal elements in a row is calculated for SLR(orange). For comparison, the same analysis on SINDy is shown in grey.
}
    
    \label{fig:SLR_general}
\end{figure}

\subsection{Effective model of center of mass acceleration through SLR}\label{acc_slr}

In this section, we model the acceleration statistics of the droplet. Unlike the droplet interface coordinates, which contain spatiotemporal data suitable for analysis via the proper orthogonal decomposition (POD) framework, the droplet acceleration is a single temporal signal extracted from direct numerical simulations (DNS) of the Cahn–Hilliard–Navier–Stokes equations. In Appendix~\ref{SINDY_acc}, we demonstrate that the SINDy framework cannot accurately model these time series, even using more advanced libraries. 
\noindent Previous studies have demonstrated that the acceleration of tracer particles in turbulent flows can be described by stochastic differential equations~\cite{PhysRevFluids.4.044301,PhysRevFluids.7.084608}. Motivated by this, we use SLR to discover analogous models that predict the acceleration statistics of the droplet. We extract the $x$ and $y$ components of the droplet acceleration, $a_x$ and $a_y$, from DNS data. Due to the homogeneity and isotropy of the system~\cite{pal2016binary}, the statistical properties of $a_x$ and $a_y$ are assumed identical. SLR is applied to $a_x$ following the approach described in Sec.~\ref{SLRinterface}, with an initial polynomial library comprising terms $[1, a_x, a_x^2, \ldots, a_x^5]$ for both the drift and stochastic components. The optimal model, which predicts acceleration statistics with minimal complexity, is identified via SSR regression. The discovered stochastic differential equation takes the form
\begin{equation}
da_x = \left(c_0 + c_1 a_x + c_3 a_x^3 \right) dt + \left(K_0 + K_1 a_x + K_2 a_x^2\right) dW_t,
\label{stoch_eqns_acc}
\end{equation}
where integration via the Euler–Maruyama scheme~\cite{kloeden1992numerical} faithfully reproduces acceleration statistics consistent with DNS results (Figure~\ref{acceleration_prediction}).
For this model (Eq.~\eqref{stoch_eqns_acc}), the cubic drift term $c_3 a_x^3$ is consistent with the Heisenberg--Yaglom prediction for Lagrangian acceleration in turbulence, where nonlinear pressure contributions generate non-Gaussian tails in the acceleration PDF~\cite{PhysRevFluids.4.044301}. The quadratic diffusion term $K_2 a_x^2$ produces state-dependent noise that amplifies at large accelerations, which is the mechanism responsible for the observed heavy-tailed (intermittent) acceleration distributions.
\noindent Previous work~\cite{pal2016binary} has shown that acceleration statistics remain invariant with respect to surface tension. In contrast, droplet size $L$ significantly affects these statistics, with intermittency intensifying as $L$ decreases. This behavior differs from the droplet interface dynamics discussed in Section~\ref{Generalisation}, where time series do not change appreciably under surface tension variations. Consequently, we expect the discovered models to capture this increase in intermittency. Consistent with these expectations, systematic variation of $L$ in our simulations reveals smooth evolution of the model coefficients\cite{PhysRevResearch.5.L042017}. Specifically, Figures~\ref{acceleration_prediction}(B,C) show continuous dependence of the drift coefficients $(c_0, c_1, c_3)$ and stochastic coefficients $(K_0, K_1, K_2)$ on $L$. The variation of stochastic coefficients with $L$ is modeled by a quadratic polynomial, whereas drift coefficients vary approximately linearly. The fitted relations are
\begin{align}
K_0 &= 2.6 \times 10^{-2} L^{2} - 7.0 L + 565.4, \\
K_1 &= -5.2 \times 10^{-2} L^{2} + 14.8 L - 1137.8, \\
K_2 &= 2.5 \times 10^{-2} L^{2} - 7.3 L + 565.2, \\
c_0 &= 1.23 L + 123.75, \\
c_1 &= -1.62 L - 95.51, \\
c_3 &= -0.33 L + 7.63.
\label{KvsLfit}
\end{align}

These results attest to the robustness of the identified model in capturing essential physical trends across a broad range of interfacial conditions.

\begin{figure}[h!] \centering \includegraphics[scale=0.143]{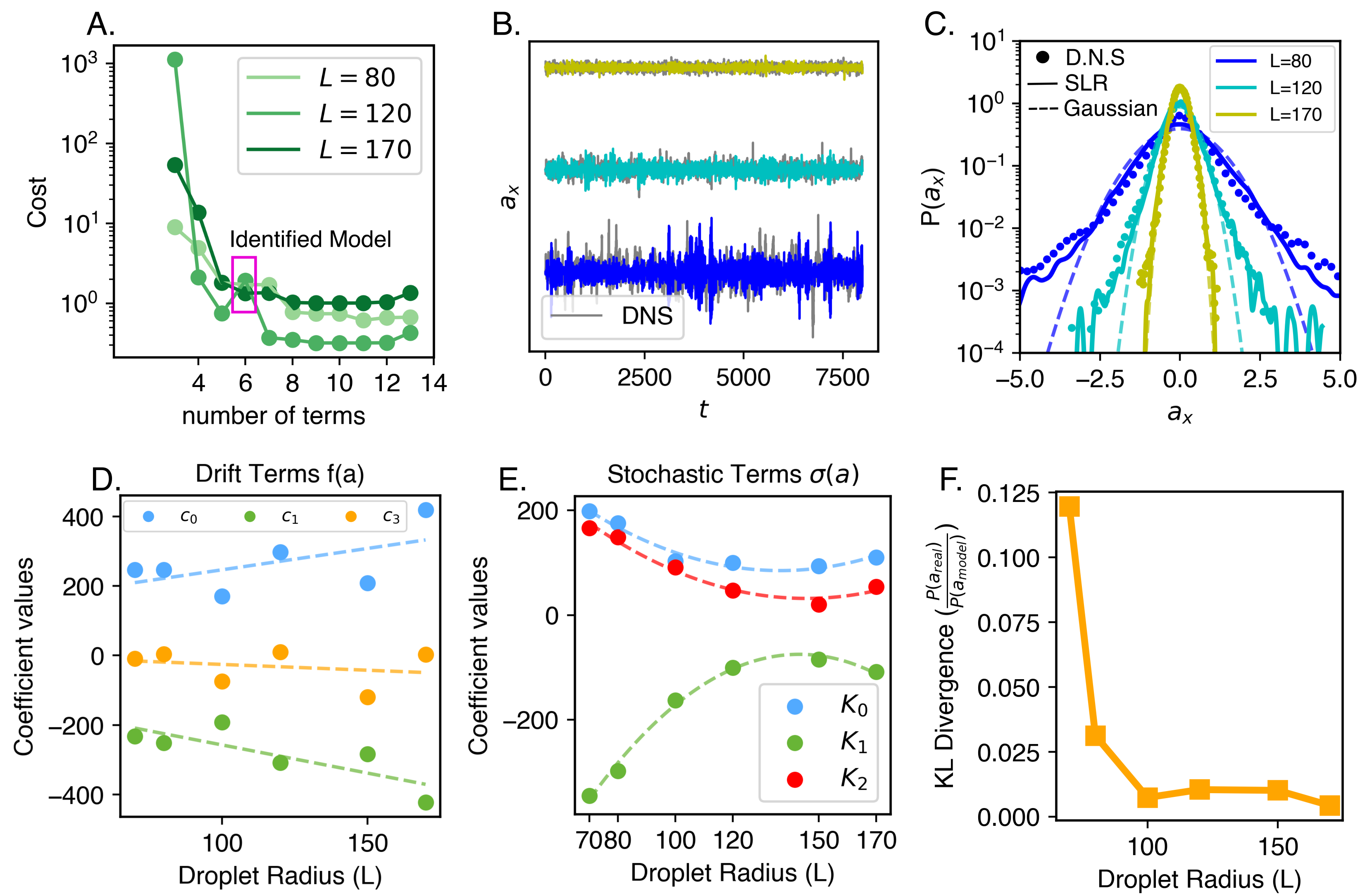} 
\caption{\textbf{Stochastic Langevin Regression for droplet acceleration:} (A) The cost function [see Eq.~\ref{cost_SLR}] at each step of sparse symbolic regression (SSR), trained on droplet data with varying initial radius $L$. During each SSR iteration, the library is initialized with 13 terms, and a term is removed until the cost function shows a noticeable increase. (B) Comparison of acceleration time series predicted by the sparse Langevin regression (SLR) model (colored curves) and direct numerical simulation (DNS) data (gray curves), with acceleration traces for each $L$ manually shifted to improve visibility and color codes the same as panel~C. (C) Probability distributions of acceleration, corresponding to panel~B: curves indicate the SLR model, dots represent DNS results, and dashed lines show the best-fit Gaussian, highlighting deviations from normality. (D) Fitted coefficients of the drift term $f(a)$ [see Eq.~\ref{stoch_eqns_acc}] as a function of droplet radius. (E) Coefficients of the stochastic term versus droplet radius. (F) KL divergence between acceleration distributions predicted by the SLR model and DNS, demonstrating that as droplet radius decreases and intermittency increases, model reliability is reduced.}

\label{acceleration_prediction}
\end{figure}

\section{Discussions and conclusion \label {conclusion}}

We compare four data-driven methodologies for identifying low-dimensional, interpretable dynamical equations: Dynamic Mode Decomposition (DMD), Hankel DMD, Sparse Identification of Nonlinear Dynamics (SINDy), and Stochastic Langevin Regression (SLR). Initial attempts to predict droplet deformation using standard DMD fail because DMD uses a linear equation to approximate the evolution of an inherently nonlinear system (Section~\ref{sec:level4}). While Hankel DMD improves performance by incorporating a time-delay embedding of past snapshots to better resolve these nonlinearities, its predictive power remains limited.

\medskip

To move beyond linear approximations, we implement the SINDy method, which leverages DNS data to discover the underlying nonlinear physical equations. As demonstrated by the KL divergence of the PDFs of the deformation parameter, SINDy significantly outperforms DMD-based approaches (see Section~\ref{sec:level4}). However, its primary limitation is poor generalization; the identified models are case-specific and fail to maintain accuracy across varying surface tensions.

\medskip

The lack of consistent predictability of the SINDy method, when we change the parameters of the system, inspired the adoption of Stochastic Langevin Regression (SLR), which couples SINDy with a stochastic framework to account for unresolved degrees of freedom (Section~\ref{acc_slr}). Unlike the deterministic DMD and SINDy methods, SLR models the system as a combination of deterministic drift and stochastic diffusion terms. Using SLR we evaluate the values of the drift and diffusion coefficients. Given the inherent randomness of turbulence, SLR provides the most reliable representation of the droplet dynamics in turbulence due to the presence of the stochastic diffusion term. We observe clear trends in drift and diffusion coefficients as a function of droplet radius (Section~\ref{acc_slr}), and also find that SLR maintains predictive accuracy across different surface tensions. SLR yields the most reliable predictive performance and the best agreement with DNS data among the four methods studied here. In addition to its accuracy, stochastic regression requires only a minimal number of modes to capture the essential dynamics in height-field fluctuations. This approach can be extended to discover stochastic evolution equations for the droplet acceleration. The model for the droplet acceleration should contain terms analogous to those derived for tracer particles in turbulent flows, along with additional, previously unreported contributions. Beyond its superior accuracy, SLR captures essential height-field fluctuations using a minimal number of modes. This framework extends to discovering stochastic equations for droplet acceleration, revealing terms analogous to tracer particles in turbulent flows alongside novel contributions. By comparing probability distributions and KL divergence between our model and DNS data, we demonstrate that physical parameters—such as surface tension and droplet size—are explicitly encoded in the identified coefficients. Our results indicate that we can obtain robust, physically interpretable dynamical equations for both the droplet height field and acceleration under turbulent forcing using data-driven techniques. This framework provides a pathway for extending such methods to other multiphysics systems where reliable reduced-order stochastic models are essential for understanding, forecasting, and controlling complex fluid–structure interactions. For example, the same framework can be applied to discover governing equations for height-field dynamics in a wide range of systems, including cell membranes, thin films, and fluid interfaces, thereby broadening its relevance well beyond the problem of droplet in turbulence studied here.

\section{Data Availability}
The code used in this study is publicly available at
\textcolor{blue}{\url{https://tinyurl.com/2uva8x58}}
. 
Due to the large size of the datasets, the raw data are available from the corresponding author upon reasonable request.
\begin{acknowledgments}
This work is supported by SERB-DST, India (SRG/2022/000163), MoE-STARS grant, and Axis Bank Center for Maths and Computing (OD/ACMC-23-0013). SZK acknowledges the MOE fellowship for financial support and Lalhminghlui for help in proofreading the manuscript. NP acknowledges support from the Faculty Start Up Grant IIT/SRIC/FSRG/2024/07. We acknowledge Nadia Bihari Padhan for help with the manuscript. 
\end{acknowledgments}

\appendix

\section{Truncated singular vector matrices  are semi-unitary}\label{SVD_s_unitary}

Given the truncated matrix \( \boldsymbol{\chi} \in \mathbb{R}^{m \times r} \) with columns \( \boldsymbol{\chi}_1, \boldsymbol{\chi}_2, \ldots, \boldsymbol{\chi}_r \), its elements are denoted as \( (\boldsymbol{\chi})_{ab} \), where \(a\) is the row index and \(b\) the column index. 

By definition of matrix multiplication, the elements of \( \boldsymbol{\chi}^T \boldsymbol{\chi} \) are:
\[
(\boldsymbol{\chi}^T \boldsymbol{\chi})_{ij} = \sum_{k=1}^{m} (\boldsymbol{\chi})_{k i} (\boldsymbol{\chi})_{k j}
\]
By construction, we know the SVD matrices are orthonormal:
\[
\boldsymbol{\chi}_i^T \boldsymbol{\chi}_j = \delta_{ij} 
\]
\noindent Thus, for diagonal elements:
\[
(\boldsymbol{\chi}^T \boldsymbol{\chi})_{ii} = \sum_{k=1}^{m} (\boldsymbol{\chi})_{k i} (\boldsymbol{\chi})_{k i} = \|\boldsymbol{\chi}_i\|^2 = 1,
\]
since each column vector has unit norm.

\noindent For off-diagonal elements (\(i \neq j\)):
\[
(\boldsymbol{\chi}^T \boldsymbol{\chi})_{ij} = \sum_{k=1}^{m} (\boldsymbol{\chi})_{k i} (\boldsymbol{\chi})_{k j} = \boldsymbol{\chi}_i^T \boldsymbol{\chi}_j = 0,
\]
due to orthogonality.
\noindent Therefore, the product matrix is the identity matrix:
\[
\boldsymbol{\chi}^T \boldsymbol{\chi} = \mathbf{I}_r,
\]
showing \( \boldsymbol{\chi} \) is semi-unitary.

\section{Verification and Validation of DMD code\label{DMD_verify}}

First, we show that our DMD model reproduces the results of a one-dimensional Burgers' equation~\cite{vergassola1994burgers,caldwell1981finite,rosenblatt1968remarks,bacshan2022nonlinear}. The Burgers' equation is a well-understood one-dimensional nonlinear equation, and serves here as a stepping stone before addressing a complex turbulent system. The general form of the 1D Burgers equation is:-
\begin{equation}
    \frac{\partial u}{\partial t} = -u\frac{\partial u}{\partial x}+\nu\frac{\partial ^2 u}{\partial ^2 x}
\end{equation}\\
where u(x,t) is the given field, $\nu$ is the diffusion coefficient (the kinematic viscosity). This is the viscous Burgers equation in 1D.
\begin{figure}[h]
\centering
\includegraphics[width=0.45\linewidth]{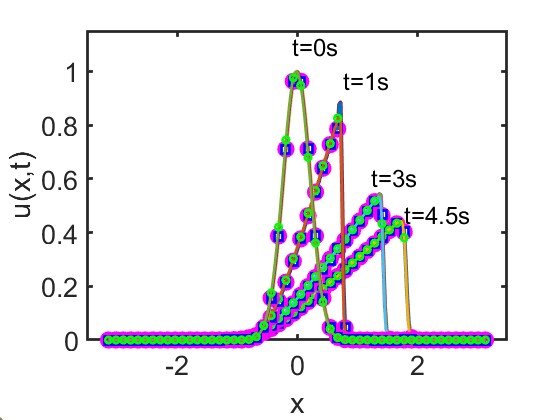}
\caption{\small The markers used in the above plots represent the following: magenta connected scatter plot(-o) is used for results from Finite Difference simulation of the 1D Burgers equation;  The blue colored square marker represents DMD (rank-reduced reconstruction) of the solution obtained from the finite difference for the same system. The cyan colored connected scatter plot represents results from the Spectral method(FFTW simulation), resolution is 256 with time steps 1000 and dt=0.005. }
\label{burgers}
\end{figure}
As Fig.~\ref{burgers} suggests, there is a reasonably good agreement between DMD and the different simulation techniques of the Burgers equation. Now we apply the same DMD method to the droplet dynamics problem.

\section{Hankel DMD}\label{HankelDMD}
Hankel Dynamic Mode Decomposition (Hankel DMD) extends the traditional DMD by incorporating delay embeddings to capture nonlinear and high-dimensional dynamics \cite{schmid2022dynamic,10.1063/5.0150689}. Given a time-series data vector \(h(t) \in \mathbb{R}^n \), we construct the Hankel matrix \(H\) as

\begin{equation}
H = 
\begin{bmatrix}
h(t_1) & h(t_2) & \cdots & h(t_m) \\
h(t_2) & h(t_3) & \cdots & h(t_{m+1}) \\
\vdots & \vdots & \ddots & \vdots \\
h(t_q) & h(t_{q+1}) & \cdots & h(t_{m+q-1})
\end{bmatrix},
\label{Hankel_matrix}
\end{equation}

where \(q\) is the embedding dimension and the total number of snapshots is \(m+q-1\).

We then form the shifted Hankel matrices

\begin{equation}
H_1 = [\, h(t_1),\, h(t_2),\, \cdots,\, h(t_{m-1}) \,], \quad
H_2 = [\, h(t_2),\, h(t_3),\, \cdots,\, h(t_{m}) \,]
\label{H1_H2_shifted}
\end{equation}

with \(H_1, H_2 \in \mathbb{R}^{qn \times (m-1)}\).

Assuming a linear operator \(A \in \mathbb{R}^{qn \times qn}\) exists such that

\begin{equation}
H_2 = A H_1 + \epsilon,
\label{A_operator_hankel}
\end{equation}

where \(\epsilon\) is the residual, we approximate \(A\) via the Singular Value Decomposition (SVD) of \(H_1\):

\begin{align}
H_1 &\approx \tilde{\chi} \tilde{\Sigma} \tilde{Q}^\dagger, \\
\tilde{\chi} &\in \mathbb{R}^{qn \times r}, \quad \tilde{\Sigma} \in \mathbb{R}^{r \times r}, \quad \tilde{Q} \in \mathbb{R}^{(m-1) \times r}, \\
A &\approx H_2 \tilde{Q} \tilde{\Sigma}^{-1} \tilde{\chi}^\dagger,
\end{align}

where \(r\) is the truncation rank.

Projecting \(A\) onto the POD basis yields the reduced operator

\begin{equation}
\tilde{A} = \tilde{\chi}^\dagger A \tilde{\chi} = \tilde{\chi}^\dagger H_2 \tilde{Q} \tilde{\Sigma}^{-1}.
\label{A_tilde_hankel}
\end{equation}

The eigenvalues \(\mu_j\) and eigenvectors \(w_j\) of \(\tilde{A}\) satisfy

\begin{equation}
\tilde{A} w_j = \mu_j w_j,
\end{equation}

where \(\mu_j\) encode temporal growth/decay rates and oscillations. The corresponding continuous-time eigenvalues are

\begin{equation}
u_j = \frac{\ln(\mu_j)}{\Delta t},
\end{equation}

and the reconstructed field is

\begin{equation}
h(\theta_i, t_k) \approx 
\sum_{j=1}^r b_j w_j(\theta_i) e^{u_j t_k},
\label{hankel_dmd_reconstruct}
\end{equation}

where coefficients \(b_j\) depend on initial conditions.

\section{Verification and Validation of the SINDY code \label{verify_SINDY}}
To assess the performance of the SINDy implementation employing the sequentially sparse regression (SSR) scheme, we benchmark the method on a known nonlinear system: the Lorenz system (Eq.~\eqref{lorentz}):
\begin{equation}
\begin{aligned}
\frac{dx}{dt} &= 10 (y - x), \\
\frac{dy}{dt} &= x (28 - z) - y, \\
\frac{dz}{dt} &= x y - \frac{8}{3} z.
\end{aligned}
\label{lorentz}
\end{equation}

The Lorenz system is numerically integrated for 13,000 temporal points with dt=0.001, with the trajectories partitioned into a training set of 11,000 points and a testing set of 2,000 points. The candidate library of observables incorporates all polynomials up to second order,
\begin{equation}
\setlength{\extrarowheight}{1ex}
\Theta = 
\left[
  \begin{array}{ccccccccc}
    \vertbar & \vertbar & \vertbar & \vertbar & \vertbar & \vertbar & \vertbar & \vertbar & \vertbar \\[-0.5ex]
    x(t) & y(t) & z(t) & x(t)y(t) & x(t)z(t) & y(t)z(t) & x(t)^2 & y(t)^2 & z(t)^2 \\[0.5ex]
    \vertbar & \vertbar & \vertbar & \vertbar & \vertbar & \vertbar & \vertbar & \vertbar & \vertbar
  \end{array}
\right].
\label{theta_def}
\end{equation}
Applying the SSR procedure iteratively until the validation error increases when more terms are removed yields the coefficient matrix $\Xi$ [Figure~\ref{sindy_test2}(A)], from which the reconstructed governing equations are obtained [Eq.~\eqref{SINDy_starting}]. The inferred model exhibits excellent quantitative agreement with the original Lorenz dynamics as shown in Figure~\ref{sindy_test2}(B). Although the exact time evolution is not reproduced, the trajectories generated by the SINDy-discovered models remain confined to the same Lorenz attractor. This behavior is a hallmark of nonlinear dynamical systems \cite{doi:10.1073/pnas.1517384113}. Consequently, if a discovered model can generate dynamics that are statistically consistent with those of the original system, even without capturing the precise trajectory, we regard it as a successful representation.
\begin{figure}[h!]
    \centering
    \includegraphics[scale=0.1]{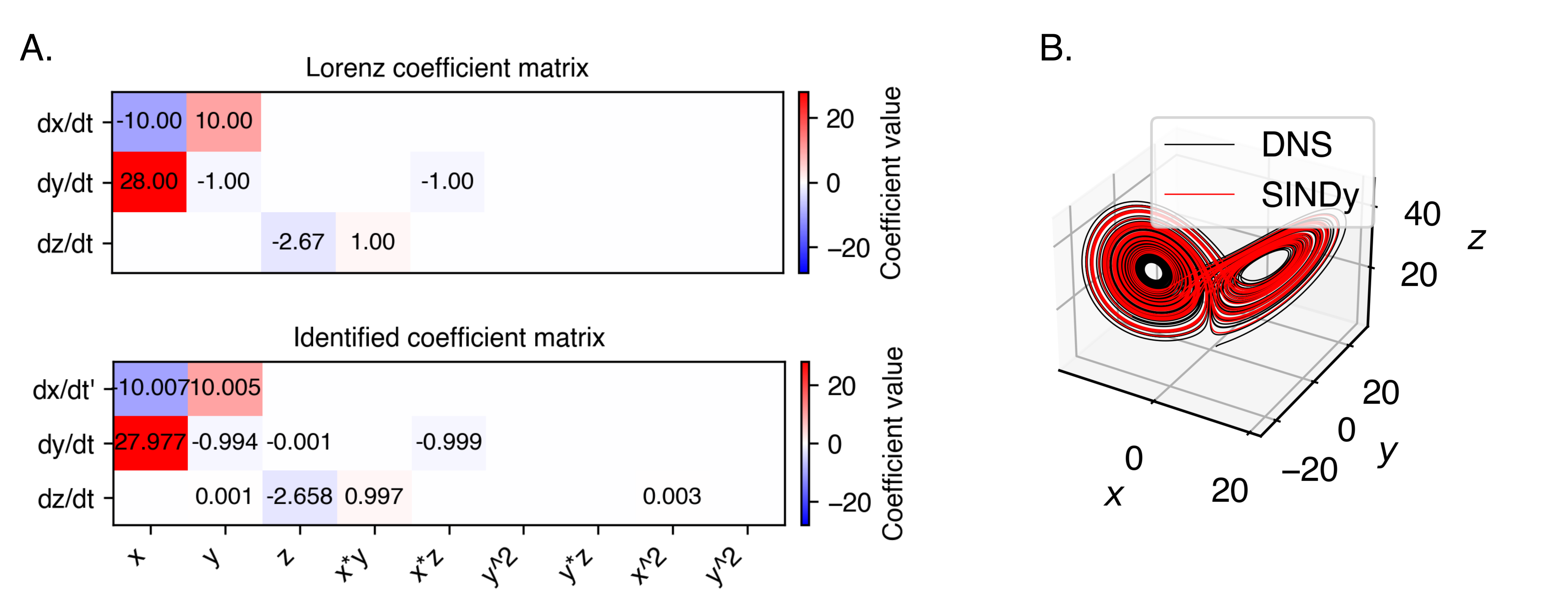}
    \caption{(A) Comparison of $\Xi$ matrices for the original system (top) and that predicted by SINDy (bottom). (B) Evolution of $x$, $y$, and $z$ over time.}
   \label{sindy_test2}
\end{figure}

\section{SINDy models fail at predicting droplet acceleration}
\label{SINDY_acc}
In Section~\ref{acc_slr}, we modeled the droplet acceleration using SLR. In this section, we show that attempts to extract governing equations using SINDy fail. The x and y components of the acceleration at each time are stored in a matrix $A \in R^{2 \times T}$ and its derivative $\dot A$ is computed. As in Section~\ref{SINDy}, we assume the dynamics follow the form
\begin{equation}
    \dot A=\Theta(A)\Xi
\end{equation}
We initialize the initial library $\Theta(A)$ using polynomial terms up to order 5, to ensure all possible terms are explored, and we also include a Fourier library with terms containing sinusoids with frequencies up to order 3. We concatenate the polynomial and Fourier libraries to include cross terms in the library. A sample of the library matrix is given below:

\begin{equation}
\Theta({A}) =
\begin{bmatrix}
1 & a_{1x}  & a_{1y} & \dots & a_{1y}^5  & \sin a_{1x} & \sin(2a_{1x}) & \dots & a_{1x}\sin a_{1x} & \dots \\
1 & a_{2x}  & a_{2y} & \dots & a_{2y}^5  & \sin a_{2x} & \sin(2a_{2x}) & \dots & a_{2x}\sin a_{2x} & \dots \\
\vdots & \vdots & \vdots & \vdots & \vdots & \vdots & \vdots & \vdots & \vdots & \vdots \\
1 & a_{T,x}  & a_{T,y} & \dots & a_{T,y}^5  & \sin a_{T,x} & \sin(2a_{T,x}) & \dots & a_{T,x}\sin a_{T,x} & \dots
\end{bmatrix}
\end{equation}
\begin{equation*}
\hspace{1.5cm}
\begin{array}{cccccccc}
\underbrace{\phantom{\quad a_{1x} \quad a_{1y} \quad a_{1x}^2 \quad a_{1y}^5}}_{\text{polynomial}} &
\underbrace{\phantom{\sin a_{1x} \quad \sin(2a_{1x}) \quad \dots}}_{\text{sinusoid}} &
\underbrace{\phantom{a_{1x}\sin a_{1x} \quad \dots}}_{\text{cross terms}}
\end{array}
\end{equation*}

We apply the sparse regression procedure outlined in Sec.~\ref{SINDy}, using an initial function library comprising a single first-order polynomial term and a single sinusoidal term of unit frequency. The discrepancy between the acceleration distribution predicted by the model and that from direct numerical simulation (DNS) is quantified via the Kullback-Leibler (KL) divergence. In Fig.~\ref{APB_sindy_acc}, we show the KL divergence for the SINDy models when the complexity of the library matrix $\Theta$ is increased by adding terms with increasing polynomial degrees and sinusoidal frequencies. We observe that even upon augmenting the library with additional terms, the KL divergence remains elevated, indicating that SINDy fails to adequately represent the acceleration dynamics even with more complex models, thus highlighting the need for the SLR method.

This limitation likely arises from the availability of solely the temporal components $a_x(t)$ and $a_y(t)$ for model construction. In contrast, interface dynamics benefit from spatiotemporal data, enabling construction of an $r$-dimensional state space via the first ten proper orthogonal decomposition (POD) modes ($q_1, \dots, q_{10}$) as shown in Section~\ref{POD_text}\cite{doi:10.2514/1.J056060}.

\begin{figure}[h!]
    \centering
    \includegraphics[scale=0.21]{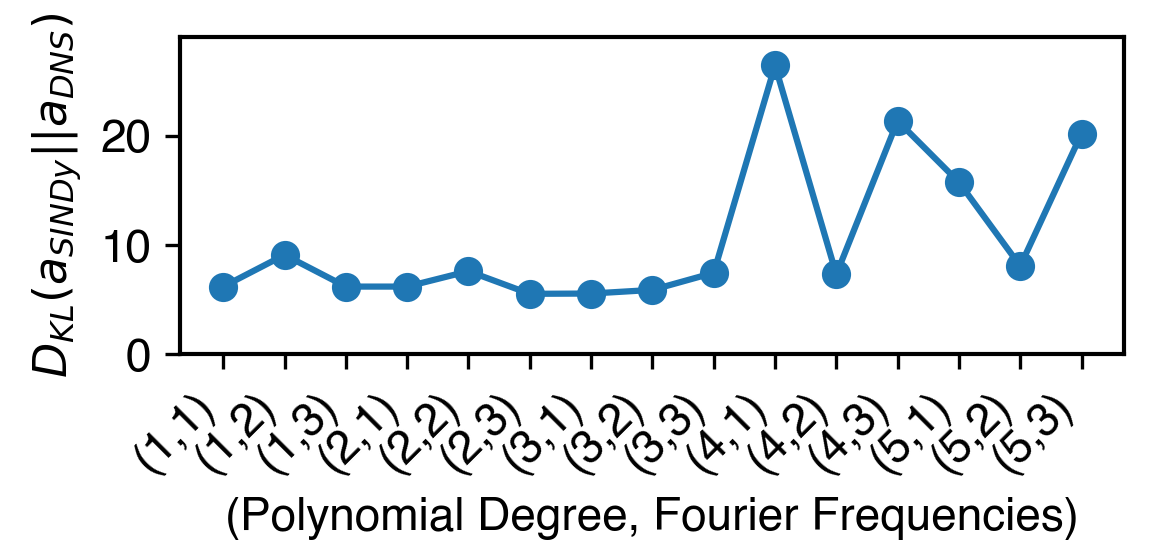}
    \caption{ Acceleration predicted from SINDy models shows that increasing the complexity by increasing the degree of the polynomial and the Fourier frequencies does not reduce the KL divergence. }
    \label{APB_sindy_acc}
\end{figure}

\section{SLR Details}\label{SLR_details}
Stochastic Langevin Regression, as mentioned in Section~\ref{SLRinterface}, is used to fit a stochastic equation to a turbulent signal and is expected to capture the intermittent features and fluctuating behaviors of turbulent signals better than SINDy, at least on a statistical level. Langevin regression is performed using the steps mentioned below.
\begin{itemize}
    \item Determining the Kramers--Moyal lag time $\tau$ is a crucial step in the success of the Langevin regression \cite{doi:10.1098/rspa.2021.0092, WANG2022244}. The determination of $\tau$ is done using two methods:
   
        One of the simplest yet most informative measures for identifying these scales is the autocorrelation function, defined as:

\begin{equation}
C(\tau) = \frac{\langle q(t + \tau) \, q(t) \rangle_t}{\langle q(t)^2 \rangle}
\end{equation}

By definition, $C(0) = 1$, and in most complex systems, $C(\tau)$ decays to zero after a certain characteristic time. This characteristic time generally reflects the macroscopic dynamics of the system. However, sampling only after the macroscopic dynamics have fully decorrelated is not desirable. Conversely, sampling when $C(\tau)$ is still close to 1 means the data remains highly correlated, which can obscure the influence of unresolved degrees of freedom.
The most insightful sampling often occurs at intermediate autocorrelation values (e.g., between 0.2 and 0.8), where fast processes cause noticeable fluctuations, but the large-scale dynamics have not yet completely lost correlation. 

Another method is by comparing the non-Markovian probability distributions with the actual probability distributions to see where the two diverge using the Kullback–Leibler (KL) divergence 
$D_{\mathrm{KL}}(\tau),$
From the definition of conditional probability, the Markov property leads to the relation:
\begin{equation}
p(q_3, t + \tau; \, q_2, t; \, q_1, t - \tau)
= p(q_3, t + \tau \mid q_2, t) \;
  p(q_2, t; \, q_1, t - \tau).
\end{equation}

This relationship can be tested directly by computing both the left-hand side and the right-hand side and checking whether they are equal for different sampling intervals $\tau$.  
  Thus, to estimate $\tau$ by comparing the two methods, we balance the requirement that the KL Divergence must be minimised and the correlation function $C(\tau)$ must be in the range 0.2--0.8. The red line drawn in Figure~\ref{APB_tau}(A) shows the chosen value of $\tau$.

\item Using the estimated $\tau$, $f_{KM}$ and $\sigma_{KM}$ are computed using Eqs.~\ref{fkm} and~\ref{akm}. The forms calculated are shown in Figure~\ref{APB_tau}(B). The functional forms of these functions are discovered using sparse regression.

\end{itemize}

A key modeling assumption in the SLR framework is that each temporal coefficient $q_j$ evolves independently. This contrasts with the SINDy formulation, where cross-coupling between modes is permitted. To assess the validity of this assumption, we computed the equal-time cross-correlation matrix $\mathcal{C}_{ij} = \langle q_i(t) q_j(t) \rangle / \sqrt{\langle q_i^2 \rangle \langle q_j^2 \rangle}$ from the DNS data. The off-diagonal elements $|\mathcal{C}_{ij}|$ for $i \neq j$ are found to be less than 0.005 for all mode pairs, indicating that the POD modes are effectively uncorrelated in the stationary state \cite{10.1098/rspa.2019.0506}.
\begin{figure}[h!]
    \centering
    \includegraphics[scale=0.13]{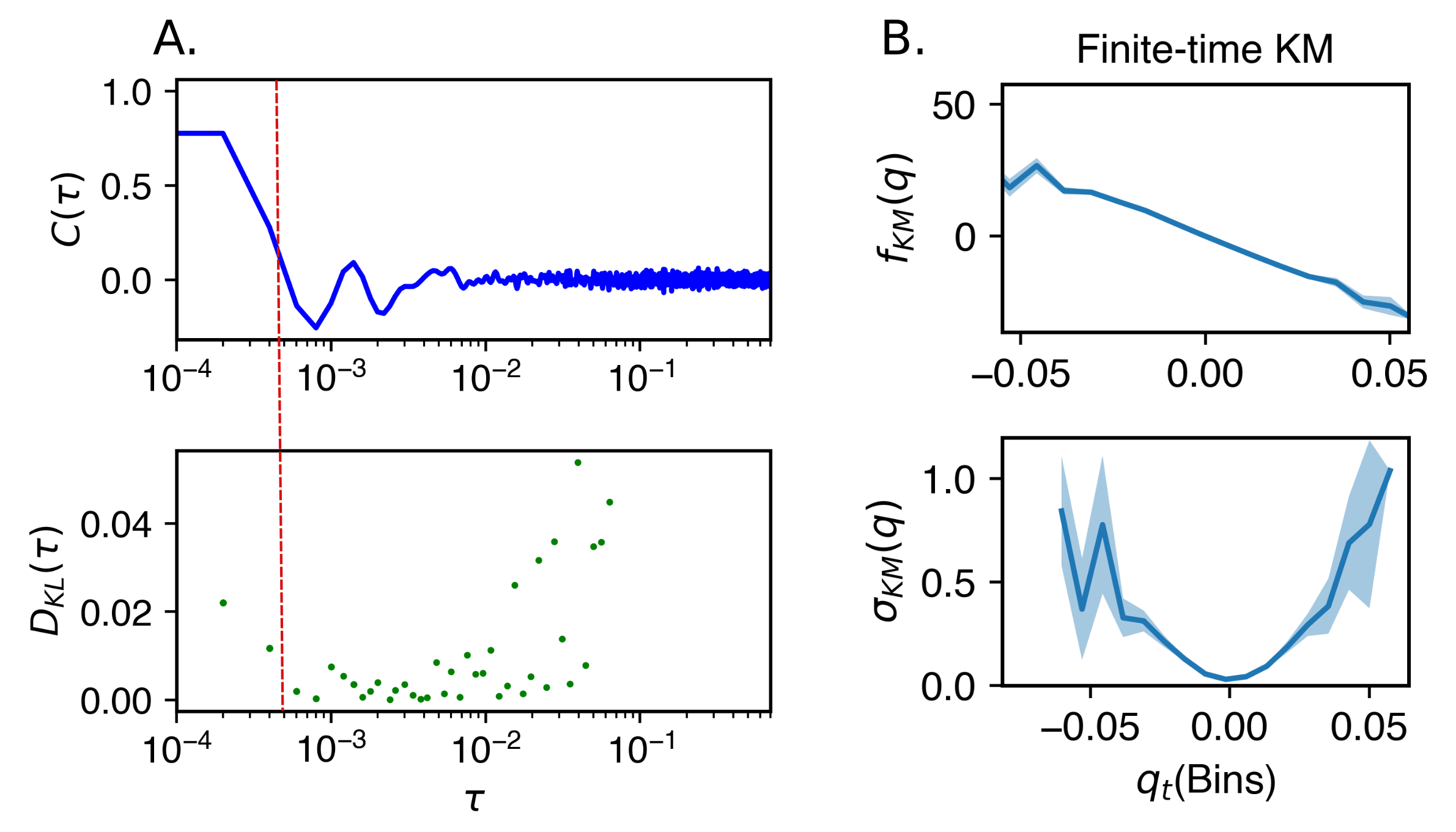}
    \caption{(A) Determination of $\tau$ for the SLR procedure by analyzing both the correlation function and KL divergence. An optimal $\tau$ is chosen that minimizes the KL divergence while keeping the correlation between 0.2--0.8.}
    \label{APB_tau}
\end{figure}

\section{Verification and Validation of the SLR code \label{verify_SLR}}
We benchmark SLR by considering a simple model of a particle moving stochastically under the influence of a double-well potential. The original Langevin equation is given below: 
\begin{equation}
dx = (20 x - x^3) \, dt + 20 \, dW_t
\label{double_well}
\end{equation}
We apply the above-mentioned SLR, as detailed in Section~\ref{SLR_text} and implemented in~\cite{doi:10.1098/rspa.2021.0092, WANG2022244}. Figure~\ref{fig:stochastic_test}(A) demonstrates that the method correctly identifies three contributing terms in the equation, consistent with the number present in Eq.~\eqref{double_well} (2 drift terms + 1 stochastic term). In Figure~\ref{fig:stochastic_test}(B), we compare the PDF of the trajectories taken from the original equation and the discovered equation and find that the PDFs of the trajectories agree.
The discovered model is $\mathrm{d}x = \big(17.54 \, x - x^3\big) \, \mathrm{d}t + 16.21 \, \mathrm{d}W_t.$

\begin{figure}[h!]
    \centering
    \includegraphics[scale=0.11]{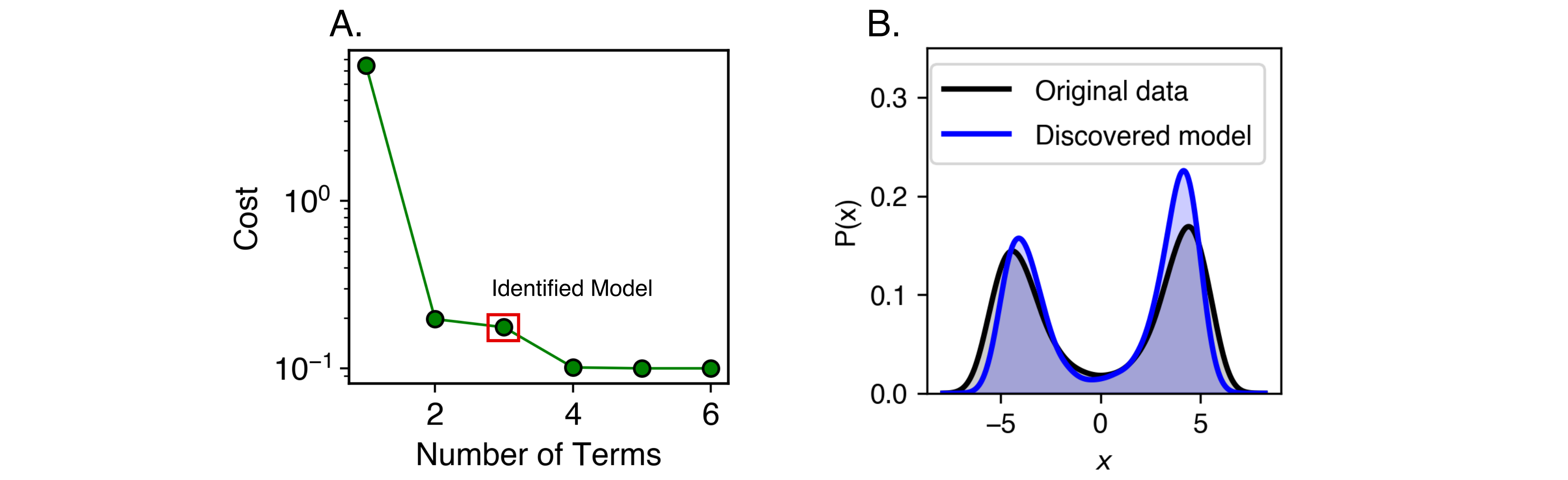}
    \caption{(A) Cost (Eq.~\ref{cost_SLR})  vs number of terms in the model at each SSR step;
    (B) PDF of the $x$ trajectory from the discovered model.
    }
    \label{fig:stochastic_test}
\end{figure}

\bibliographystyle{apsrev4-2}
\bibliography{apssamp}

\end{document}